\documentclass[12pt]{iopart}
\usepackage{graphicx,iopams}
\begin{document}


\title[Influence of local restrictions on stochastic predator--prey models]
        {Influence of local carrying capacity restrictions on 
	stochastic predator--prey models}

\author{Mark J Washenberger$^1$, Mauro Mobilia$^2$, and Uwe C T\"auber$^1$}

\address{$^1$ Department of Physics and 
	Center for Stochastic Processes in Science and Engineering,  
	Virginia Polytechnic Institute and State University, \\
	Blacksburg, VA 24061--0435, USA}

\address{$^2$ Arnold Sommerfeld Center for Theoretical Physics and 
	Center for NanoScience, Department of Physics,
	Ludwig--Maximilians--Universit\"at M\"unchen, \\
  	D-80333 Munich, Germany}

\eads{\mailto{mwashenb@vt.edu}, \mailto{mauro.mobilia@physik.lmu.de}, 
	\mailto{tauber@vt.edu}}

\begin{abstract}
We study a stochastic lattice predator--prey system by means of Monte Carlo
simulations that do not impose any restrictions on the number of particles per
site, and discuss the similarities and differences of our results with those 
obtained for site-restricted model variants.
In accord with the classic Lotka--Volterra mean-field description, both species
always coexist in two dimensions.
Yet competing activity fronts generate complex, correlated spatio-temporal 
structures.
As a consequence, finite systems display transient erratic population 
oscillations with characteristic frequencies that are renormalized by 
fluctuations.
For large reaction rates, when the processes are rendered more local, these 
oscillations are suppressed. 
In contrast with site-restricted predator--prey model, we observe species 
coexistence also in one dimension.
In addition, we report results on the steady-state prey age distribution.
\end{abstract}
\submitto{\JPCM --- \today}


\section{Introduction}

Originally devised to describe autocatalytic chemical reactions \cite{Lotka}
and fish harvests in the Adriatic \cite{Volterra}, the classic Lotka--Volterra
coupled set of ordinary differential equations represents a central paradigm
for the emergence of species coexistence and periodic oscillations in nonlinear
systems with competing constituents. 
It thus features prominently in textbooks on nonlinear dynamics \cite{Haken}, 
ecology \cite{May, Maynard}, population dynamics \cite{Sigmund,Neal}, and 
mathematical biology \cite{Murray}, although this {\em deterministic} rate
equation system is known to be mathematically unstable against modifications,
spatial variations, and stochasticity, and therefore also unlikely to be 
biologically relevant \cite{Neal, Murray}.

Neither criticism however pertains to {\em stochastic} spatial predator--prey 
models \cite{Matsuda, Tome, Boccara, Durrett, Provata, Albano, Lipowski, 
Lipowska, Monetti, Shnerb, Droz, Antal, Kowalik, McKane, Georgiev, Mobilia}, 
for which the mean-field approximation recovers the original Lotka--Volterra 
differential equations.
In stark contrast with the deterministic Lotka--Volterra rate equations, such
stochastic lattice predator--prey models in fact display remarkably robust 
features (for a recent overview, see \cite{Mobilia}):
sufficiently deep in the species coexistence phase, the population densities 
oscillate in an irregular manner, however with characteristic periods and 
amplitudes that vanish in the thermodynamic limit \cite{Provata, Lipowski, 
Lipowska, Monetti, Antal, Kowalik, Georgiev, Mobilia}; these erratic 
oscillations are induced by recurrent activity waves that initially form 
concentric rings, and upon merging produce complex spatiotemporal structures 
\cite{Boccara, Droz, Georgiev, Mobilia}.

These quite generic characteristics are found in computer simulations of 
several distinct model variants, differing in the precise microscopic 
algorithmic setups such as the number of possible states per lattice site and 
detailed implementations of the reaction scheme, e.g., parallel vs. sequential
updates, etc.
Perhaps suprisingly this includes even long-range processes \cite{Lipowska, 
Monetti}, generalizations to `smart' predators and prey who respectively follow
/ evade the other species \cite{Boccara, Albano}, and a variation where the 
predation reaction is split up into two separate and independent processes
\cite{Georgiev}.
Indeed, already a zero-dimensional stochastic predator--prey model exhibits
prominent population oscillations driven by the inevitable internal reaction
noise \cite{McKane}; such finite-size fluctuations also drastically affect the
properties of, e.g., the three-species cyclic Lotka--Volterra model
\cite{Reichenbach}. 

Largely for computational simplicity, such lattice models for reacting particle
systems typically operate with constraints on the possible site occupation
numbers, usually allowing only at most one particle per site; but four-state 
predator--prey systems which permit both a single predator and prey per site
have been studied too \cite{Lipowski, Lipowska, Antal}.
Biologically and ecologically, such site occupation number restrictions may be 
interpreted as originating in {\em local} limitations on resources for either
species.
On the mean-field description level, they are represented by finite carrying
capacities in the rate equations that prevent unbounded (Malthusian) species 
growth (see, e.g., \cite{Neal, Murray}).
Already on this rate equation level, such constraints have a dramatic effect:
they lead to the emergence of an extinction threshold for the predator
population.
In the stochastic spatial system (and in the thermodynamic limit), predator 
extinction is governed by a continuous active-to-absorbing phase
transition.
As one would expect \cite{Hinrichsen, Odor, Janssen, Howard}, the associated 
critical exponents are those of directed percolation \cite{Tome, Boccara, 
Albano, Lipowski, Lipowska, Monetti, Antal, Kowalik}; an analytic argument that
demonstrates this assertion, based on a field theory representation of the
underlying stochastic processes (see, e.g., \cite{Howard}) is presented in
\cite{Mobilia}.

Presumably already a lattice representation should be viewed as a mesoscopic 
representation of the chemical or biological system under consideration; 
naturally therefore the question arises which influence the implicit 
coarse-graining scale might have on the system's properties.
Also, it is of interest to study a stochastic spatial system most closely
related to the original Lotka--Volterra model, i.e., without any limits on the
carrying capacities (specifically for the prey).
In this work, we investigate such an unconstrained lattice predator--prey 
model.
As suggested by the corresponding mean-field theory, we do not encounter 
predator extinction, but both species always coexist (within the typical time
scales of our simulation runs).
This appears to be true even in one dimension.
The strictly periodic population oscillations, which are moreover determined by
the initial configuration, of the deterministic rate equations are replaced 
with erratic, transient oscillations largely determined by the intrinsic rates,
and with frequencies renormalized by the fluctuations.
If the reaction rates are high, and the processes effectively rendered more
local, we find the oscillatory behaviour to cease.
As in the lattice models with site occupation number restrictions, persistent
complex spatio-temporal structures form, but the activity fronts are generally 
more diffuse.

In the following section, we briefly review the results of the mean-field rate
equation approximation for Lotka--Volterra type predator-prey models with and
without limiting carrying capacities.
Next we describe the alterations in the Monte Carlo algorithm and data 
structure that are required if we wish to allow arbitrarily many occupants per 
lattice site, before we present our simulation results.
Finally, we summarize and discuss our findings in the concluding section.

\section{Model and mean-field rate equations}

\subsection{Unconstrained Lotka--Volterra rate equations}

We wish to consider a two-species system of diffusing particles or population
members subject to the following reactions:
\begin{eqnarray}
  &A \to \emptyset \ , \quad &{\rm rate} \ \mu \ , \nonumber \\
  &A + B \to A + A \ , \quad &{\rm rate} \ \lambda \ , \\
  &B \to B + B \ , \quad &{\rm rate} \ \sigma \ . \nonumber
\label{lvreac}
\end{eqnarray}
The `predators' $A$ die spontaneously at rate $\mu > 0$, whereas the `prey' $B$
proliferate with rate $\sigma > 0$.
In the absence of the binary `predation' interaction with rate $\lambda$, the 
uncoupled first-order processes would naturally lead to predator extinction 
$a(t) = a(0) \, \rme^{- \mu t}$, and Malthusian prey population explosion
$b(t) = b(0) \, \rme^{\sigma t}$.
Here, $a(t)$ and $b(t)$ respectively denote the $A$/$B$ concentrations or 
predator/prey population densities.
The second reaction in the above scheme \eref{lvreac} induces species 
coexistence \cite{Lotka, Volterra}.

All three reactions \eref{lvreac} as well as nearest--neighbour hopping are to
be interpreted as {\em stochastic} processes, and the spatial distribution of
reactants as well as reaction-induced correlations turn out to be relevant for 
a quantitative characterization of the kinetics of this system.
Nevertheless, straightforward mean-field theory provides relevant insight into
some basic properties of the system, and are reflected in the full stochastic
simulation results to be discussed in section~4 below.
Thus let us first ignore spatial variations, fluctuations and correlations, 
which leads to the corresponding {\em mean-field} rate equations for the 
average concentrations:
\begin{equation}
  \dot{a}(t) = \lambda \, a(t) \, b(t) - \mu \, a(t) \ , \qquad
  \dot{b}(t) = \sigma \, b(t) - \lambda \, a(t) \, b(t) \ .
\label{lvreqa}
\end{equation}
Setting the time derivatives to zero yields three stationary states 
$(a_s,b_s)$: (i) the absorbing state with total population extinction $(0,0)$, 
which is obviously linearly unstable (if $\sigma > 0$); 
(ii) predator extinction and prey explosion $(0,\infty)$, which for 
$\lambda > 0$ is also linearly unstable (and represents an absorbing state for 
the predators); 
and (iii) species coexistence $(a_u = \sigma/\lambda , b_u = \mu/\lambda)$.
This fixed point is however only marginally stable, for the eigenvalues of the
Jacobian stability matrix are purely imaginary, $\rmi \sqrt{\mu \, \sigma}$.
Indeed, linearizing \eref{lvreqa} near $(a_u,b_u)$ results in the coupled 
differential equations $\delta \dot{a}(t) = \sigma \, \delta b(t)$,
$\delta \dot{b}(t) = - \mu \, \delta a(t)$, which are readily solved by 
$\delta a(t) = \delta a(0) \, \cos\left( \sqrt{\mu \, \sigma} \, t \right) + 
\delta b(0) \, \sqrt{\sigma / \mu} \, \sin\left( \sqrt{\mu \, \sigma} \, t 
\right)$ and $\delta b(t) = - \delta a(0) \, \sqrt{\mu / \sigma} \, 
\sin\left( \sqrt{\mu \, \sigma} \, t \right) + \delta b(0) \, 
\cos\left( \sqrt{\mu \, \sigma} \, t \right)$.

\begin{figure}[t] \begin{center}
\includegraphics[width = 7.7cm]{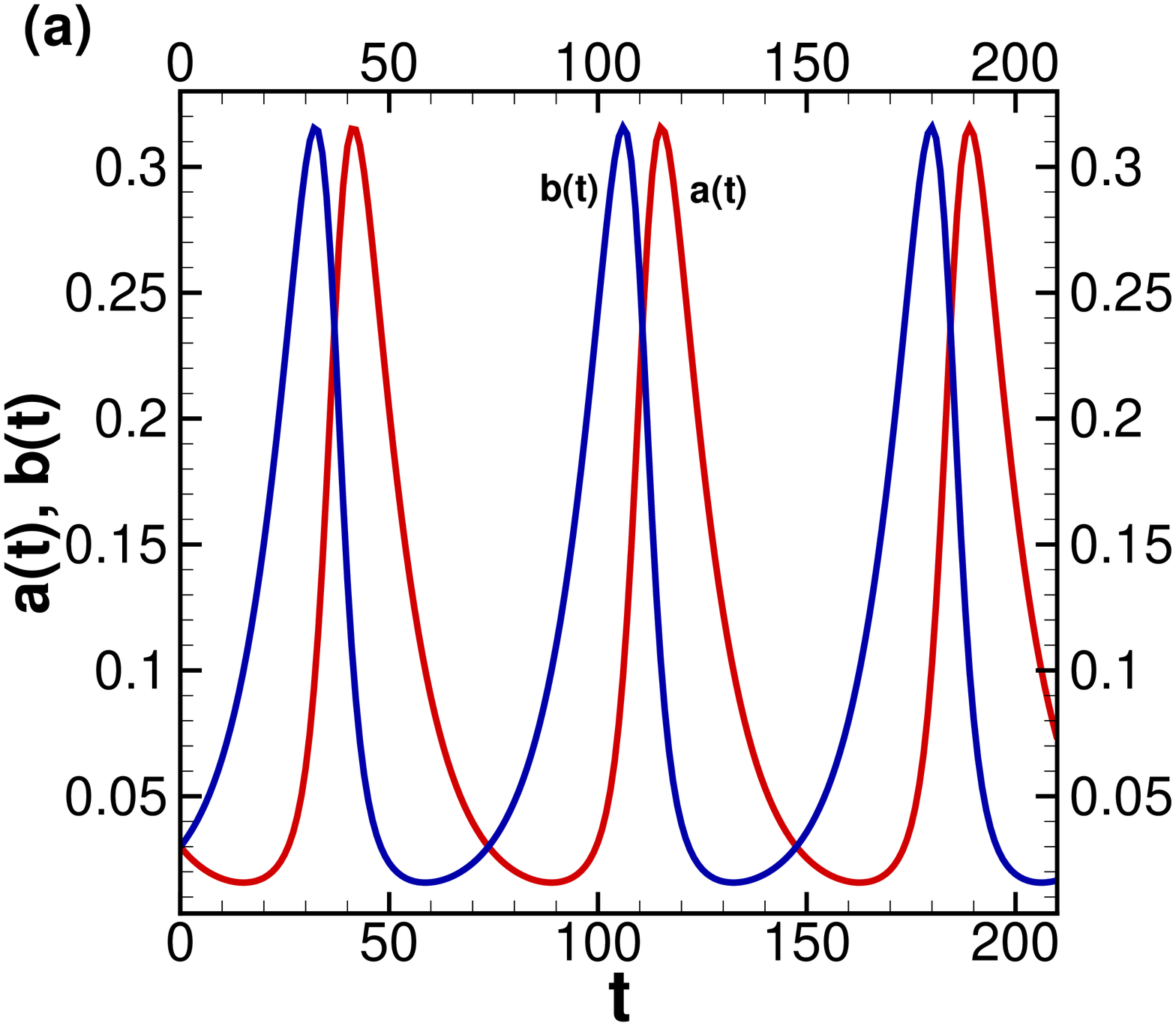} \
\includegraphics[width = 7.7cm]{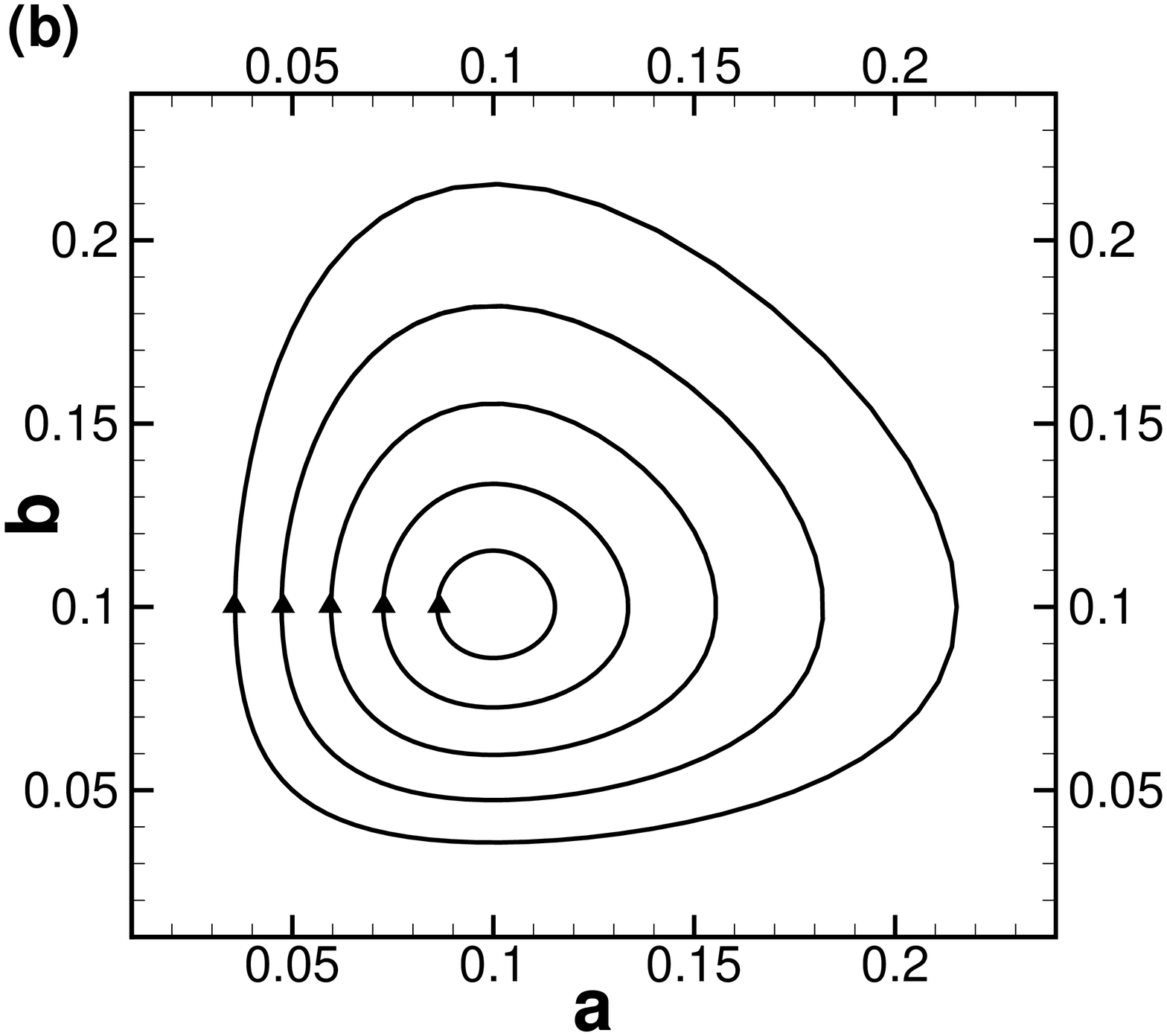}
\caption{\label{mforbt}
	(a) Predator (red) and prey (blue) population oscillations resulting 
	from the deterministic Lotka--Volterra equations \eref{lvreqa}, all 
	computed with rates $\sigma = 0.1$, $\mu = 0.1$, and $\lambda = 1$.
	(b) Several periodic orbits in the $a$-$b$ phase plane. The oscillatory
	kinetics is determined by the initial conditions. (Colour online.)}
\end{center}
\end{figure}

This suggests periodic oscillations about the center fixed point $(a_u,b_u)$
with frequency $f = \sqrt{\mu \, \sigma} / 2 \pi$. 
Indeed, solving for the phase space trajectories yields $\rmd a / \rmd b = 
[a \, (\lambda \, b - \mu)] / [b \, (\sigma - \lambda \, a)]$, with a first 
integral (equal to the negative expression of the associated Lyapunov function 
for the rate equations with growth-limiting term, see below)
\begin{equation}
  K(t) = \lambda [a(t) + b(t)] - \sigma \ln a(t) - \mu \ln b(t)
\label{mfcons}
\end{equation}
that is conserved under the kinetics \eref{lvreqa},
$K(t) = K(0)$.
The solutions of the {\em deterministic} mean-field Lotka--Volterra model are
thus closed orbits in phase space, i.e., regular periodic nonlinear population 
oscillations whose amplitudes and frequencies determine crucially on the
{\em initial} values $a(0)$ and $b(0)$, as depicted in figure~\ref{mforbt}.
Notice however that there is no physical symmetry behind the conservation law 
\eref{mfcons}; $K(t) = {\rm const.}$ is a mere mathematical property of the
coupled mean-field rate equations, and will not hold for the underlying 
stochastic process.

\subsection{Predator--prey rate equations with limited prey carrying capacity}

These regular population oscillations fixed by the initial state are clearly 
not realistic in a biological setting.
The Lotka--Volterra system is therefore often rendered `more appropriate' 
through introducing a growth-limiting term for the prey \cite{Neal, Murray}, 
whence the second differential equation in \eref{lvreqa} is replaced with
\begin{equation}
  \dot{b}(t) = \sigma \, b(t) \left[ 1 - \rho^{-1} \, b(t) \right] 
  - \lambda \, a(t) \, b(t) \ . 
\label{lvreqc}
\end{equation}
The new parameter $\rho > 0$ can be interpreted as the prey {\em carrying 
capacity} (maximum total population density).
We remark in parentheses that the rate equations \eref{lvreqa} and 
\eref{lvreqc} may be derived in a systematic manner from the underlying 
stochastic master equation for the stochastic processes \eref{lvreac} with site
occupation numbers restricted to $0$ or $1$ for either species \cite{Mobilia}.

The non-trivial stationary states in the ensuing restricted Lotka--Volterra 
model become modified to (ii) predator extinction and prey saturation 
$(0,\rho)$, linearly stable for $\lambda < \lambda_c = \mu / \rho$; 
and (iii) species coexistence $(a_r,b_r)$ with $b_r = \mu / \lambda$ and
$a_r = (\sigma / \lambda) \, (1 - \mu / \lambda \rho)$, which requires 
efficient predation, $\lambda > \lambda_c$.
In this case the coexistence fixed point is always linearly stable, since the
associated eigenvalues of the Jacobian,
$\epsilon_\pm = - \sigma \mu (2 \lambda \rho)^{-1} \left[ 1 \pm \sqrt{1 - 
4 \lambda \rho \sigma^{-1} \left( \lambda \rho \mu^{-1} - 1 \right)} \right]$,
have a negative real part indicating exponential approach to $(a_r,b_r)$.
For $\sigma > \sigma_s = 4 \lambda \, \rho \left( \lambda \rho / \mu - 1 
\right) > 0$, or $\mu / \rho < \lambda < \lambda_s = \left( 1 + 
\sqrt{1 + \sigma/\mu} \right) \mu / 2 \rho$, these eigenvalues are real, and
the fixed point can be characterized as a stable node.
On the other hand, if $\sigma < \sigma_s$ or $\lambda > \lambda_s$, i.e., deep 
in the coexistence phase, $\epsilon_\pm$ turn into a complex conjugate pair, 
and $(a_r,b_r)$ becomes a stable spiral singularity which is approached in a 
damped oscillatory manner. 
Actually, going beyond the linear analysis, the existence of the Lyapunov 
function ${\cal L} = \lambda [ a_r \ln a(t) + b_r \ln b(t) - a(t) - b(t)]$ for 
the rate equations \eref{lvreqa} and \eref{lvreqc} implies the global stability
of the reactive fixed point $(a_r,b_r)$ \cite{Sigmund,Murray}.

Thus already within the mean-field approximation a finite prey carrying 
capacity $\rho$, which can be viewed as the average result of local 
restrictions on the $B$ density, drastically affects the phase diagram by 
inducing an extinction threshold (at $\lambda_c$ for fixed $\mu$) for the 
predator population.
Taking spatial fluctuations into account, this becomes a genuine continuous 
active-to-absorbing phase transition for the $A$ species. 
It is now well-established that its critical properties are governed by the 
scaling exponents of directed percolation, with critical dimension $d_c = 4$ 
(see \cite{Mobilia} and references therein).

\section{Monte Carlo Simulations}

\subsection{Data Structure}

Monte Carlo simulations of chemical kinetics on a lattice are usually 
performed with site occupation number restrictions (e.g., any lattice site may
be occupied by at most one particle of either species) since this situation is
readily coded in a straightforward and efficient manner.
Indeed, for any stochastic simulation of chemical kinetics, a lack of site 
restrictions presents a challenge for efficient algorithms and data storage. 
Unlike the restricted case, no upper bound can be placed on the number of 
occupants of a lattice site. 
Thus, the memory structure must be made to follow the particles interacting on
the lattice rather than just the lattice itself. 
This requirement calls for a more dynamic implementation of the simulation than
simply flipping bits on a two-dimensional array.

In this subsection, we briefly overview several approaches to storing the 
particle information, and compare them in light of a decomposition of the 
demands the simulation will place on the data structure. 
We then present and explain the hybrid data structure utilized in our 
simulations.
In essence, simulating diffusing and interacting particles on a lattice 
requires four operations repeated in different combinations, namely
\begin{enumerate}
\item Random selection: a lattice occupant is selected at random with a 
	uniform probability, which allows for the simulation to proceed in an 
	unbiased manner.
\item Query: this operation determines the number of particles of any species
	located on a given site at a specific instant; it is desirable to 
	render this operation fast which puts some constraints on the particle
	ordering in the data structure.
\item Add: a new particle is inserted into the lattice.
\item Remove: annihilates a particle from the lattice; a particle movement can
	be implemented as sequential Remove and Add.
\end{enumerate}
Considering these operations provides us with a frame in which to compare the 
competing data structures and algorithms for implementing site-unrestricted 
stochastic lattice simulations for particle reactions.

The first such candidate, a static array, is perhaps the most intuitive. 
In this implementation, we simply construct arrays (in arbitrary $d$ 
dimensions) of integer $n$-tuples, where each integer tells how many particles 
of a given species currently reside on that site. 
The problem of unbound variables is avoided by imposing some loose site 
restrictions, say by setting a cap of some large maximum number $M$ of 
particles per site. 
Such a restriction would be irrelevant to the chemical kinetics unless we 
actually encounter such a large number on a single site, which is typically not
likely. 
However, this implementation presents major problems with random selection. 
When restriction is in place, especially strong restrictions, such as allowing
only a single occupant per site, the typical algorithm proceeds by selecting
sites at random repeatedly until an occupied one is found. 
This can be duplicated for any site restriction, including very loose ones. 
However, with $L^d$ sites, $N$ occupants, and a site restriction of $M$ 
particles per site, this algorithm has a time complexity of $O(L^d M / N)$, 
and thus very inefficient for very large $M$.
A more reliable algorithm, albeit still relatively slow, involves a full 
traversal of the lattice; therefore, the best we can do here is $O(L^d)$.

The next possible candidate is an unordered list with random access. 
This implementation would simply construct a very long list at the beginning of
the simulation, and then proceed to populate it with the lattice occupants, 
where each entry keeps track of its own location in the lattice. 
This structure gives very fast random selection, insertion, and removal, all
$O(1)$. 
However, since it is unordered, a query could easily require searching the 
entire list, which results in an $O(N)$ performance for the query.
We might try to enhance the query performance by ordering the list or 
constructing it as a tree; in either case the search efficiency is improved to
$O(\log N)$. 
But we also degrade the insertion and removal to $O(\log N)$. 
In the process, if we have constructed these data structures within the 
confines of an array so that we can maintain random access, random selection 
remains sufficiently fast, at $O(1)$. 
However, if we lose random access, as in the traditional linked list or binary 
search tree, we arrive at $O(N)$ complexity for the random selection, since any
random selection would require an ordered traversal of the tree or list.

Therefore, the data structure we propose hybridizes the first two candidates
above to ensure constant time complexity for each of the necessary operations. 
What is required is to use a spatial array to allow fast addition, removal and 
queries, but to manage the memory in such a way as to provide random access to
an unordered list of the occupants of the sites, therefore enabling immediate 
access for the uniformly random selection of an occupant. 
To accomplish this hybridization, we keep an unordered list of all the current 
occupants of the lattice. 
Each of these maintains information about where in the lattice it is located. 
In turn, we maintain a spatial array of the lattice, where each site contains a
pointer to the head of a linked list of occupants. 
The fact that these occupants are stored within the global, unordered list, 
imposes no constraint on our ability to render them elements of local linked 
lists.
If for a given simulation it is desired to maintain unique information about 
each individual lattice occupant apart from its location, then this 
implementation does require some traversal of the site-local linked lists. 
This can become quite burdensome as the expected occupancy of a site grows. 
Yet if we can sacrifice such individual information, we need only ever interact
with the head of each site-local list, or an occupant that we have already 
found directly through the random selection. 
Thus, constant time complexity can be maintained.

\subsection{Monte Carlo Simulation Procedure}

For each iteration of the simulation, a lattice occupant (either predator or 
prey) is selected at random, and hops to a nearest-neighbour site, and may
subsequently be subjected to an (on-site) reaction. 
To keep track of the evolution of the system through time $t$, each such step 
is accompanied by an increment in time equal to $1/N(t)$, where $N(t)$ is the 
total number of occupants at this instant. 
In this way, in a single time step, on average every occupant has been once
selected for interaction.

Our specific Monte Carlo simulation for the unconstrained stochastic lattice
Lotka--Volterra model proceeds as follows:
\begin{enumerate}
\item Select a lattice occupant at random.
\item Select one of the $2d$ sites (on a hypercubic lattice in $d$ dimensions)
	adjacent to this occupant, and move it there (nearest-neighbour 
	hopping).
\item If the occupant is a $B$ particle (prey): \\
 	generate a random number $r_1$ over the range $[0,1)$; 
 	if $r_1 < \sigma$, add another new $B$ particle to the current site 
	(prey offspring production, $B \to B + B$).
\item If the occupant is an $A$ particle (predator):
\begin{enumerate}
\item if there are any $B$ particles (prey) located on this site:
	for {\em each} $B$ particle generate a random number $r_2$ over the 
	range $[0,1)$; if $r_2 < \lambda$, remove it from and add one new $A$ 
	particle to this site (predation interaction, $A + B \to A + A$);
\item generate a random number $r_3$ over the range $[0,1)$; 
	if $r_3 < \mu$, remove this $A$ particle 
	(predator death, $A \to \emptyset$).
\end{enumerate}
\end{enumerate}

In our simulations, none of the possible events are mutually exclusive; in 
particular, particle diffusion and on-site reactions can happen simultaneously.
Therefore, the hopping probability is held fixed at $1$ for the entire range of
reaction parameters, which obviates the need for renormalization of the 
simulation time for different reaction rates.
Notice that in contrast with site-restricted simulations, where offspring
particles need to appear at neighbouring sites which causes either species to
propagate in space, here hopping processes have to be put in explicitly.
All simulation runs reported below start from random initial spatial 
distributions of both predators $A$ and prey $B$.
We have also run simulations where step (iv a) above was modified:
instead of exposing each prey on a given site to predation by the invading
predator, we allowed only at most a single $B$ particle to be removed.

\section{Simulation results}

\subsection{Spatio-temporal structures and spatial correlations in two 
            dimensions}

\begin{figure}[t] \begin{center}
\includegraphics[width = 15.7cm]{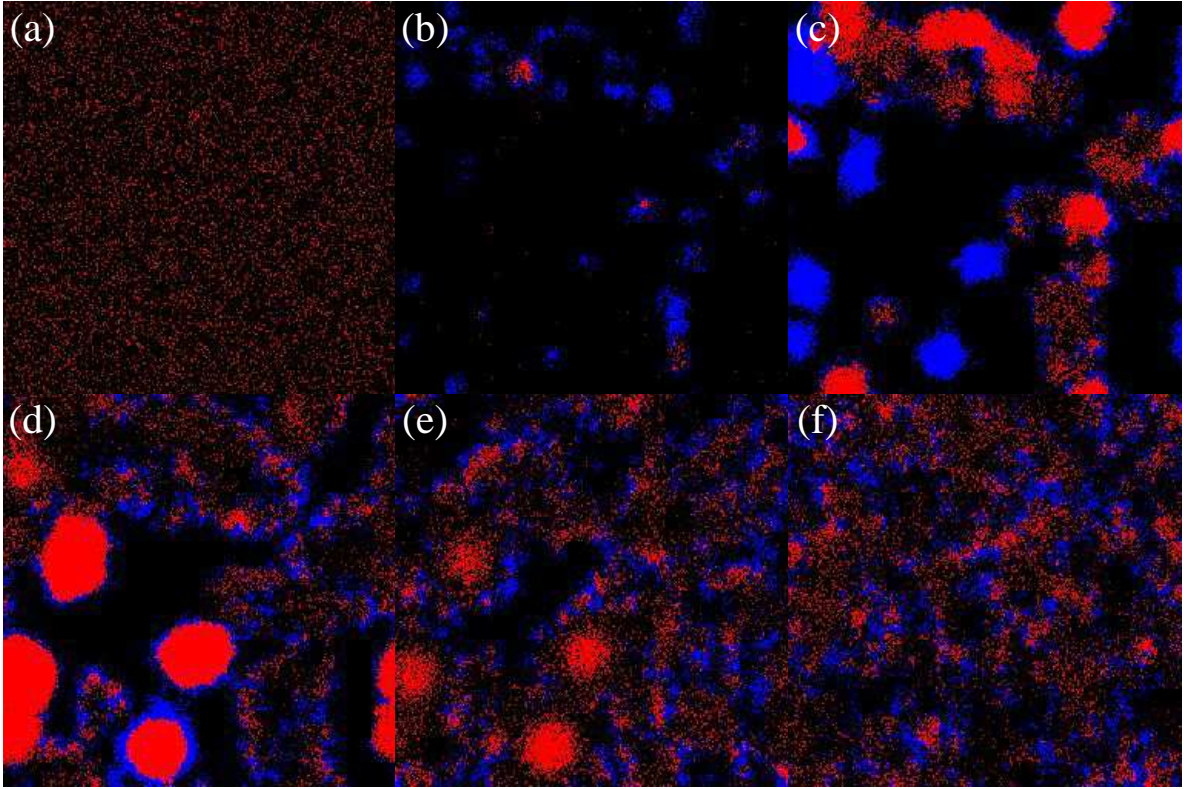}
\caption{\label{snapsh}
	Snapshots taken from a simulation on a $256 \times 256$ lattice, with
	rates $\sigma = 0.1$, $\mu = 0.1$, and $\lambda = 1.0$. 
	Red and blue pixels respectively indicate sites with at least one 
	predator ($A$) and prey ($B$); magenta sites are occupied by at least
	one representative of either species; black sites are empty.
	The simulation proceeds from (a) to (f), with pictures taken at time 
	$t = 0$, $50$, $95$, $129$, $170$, and $600$. (Colour online.)}
\end{center}
\end{figure}

Figure \ref{snapsh} depicts a set of snapshots from a simulation on a 
$256 \times 256$ square lattice, run with rate parameters $\sigma = 0.1$, 
$\mu = 0.1$, and $\lambda = 0.1$ \cite{Movies}. 
Starting from a random (Poisson) distribution of particles $A$ and $B$ (a), 
initially the predator population goes almost extinct, with a few localized 
specks of prey surviving.
When predators encounter the regions inhabitated by the prey, they rapidly
devour them, and subsequently die out (c,d,e).
Eventually, the system settles in a dynamic steady state governed by expanding,
competing, and merging diffuse activity fronts, forming complex spatio-temporal
structures (f).
This temporal evolution resembles that observed in simulations with restricted 
site occupation numbers deep in the species coexistence phase (compare, e.g., 
figure~2 in \cite{Mobilia} and \cite{Movies}), except that the fronts are more 
diffuse in the unrestricted simulations, broadened by regions that contain both
predators and prey (colour-coded magenta / light grey).
Yet there is also a marked difference, namely the predator--prey competition is
far more local and thus considerably faster in the present simulation runs.

\begin{figure}[t] \begin{center}
\includegraphics[width = 5.1cm]{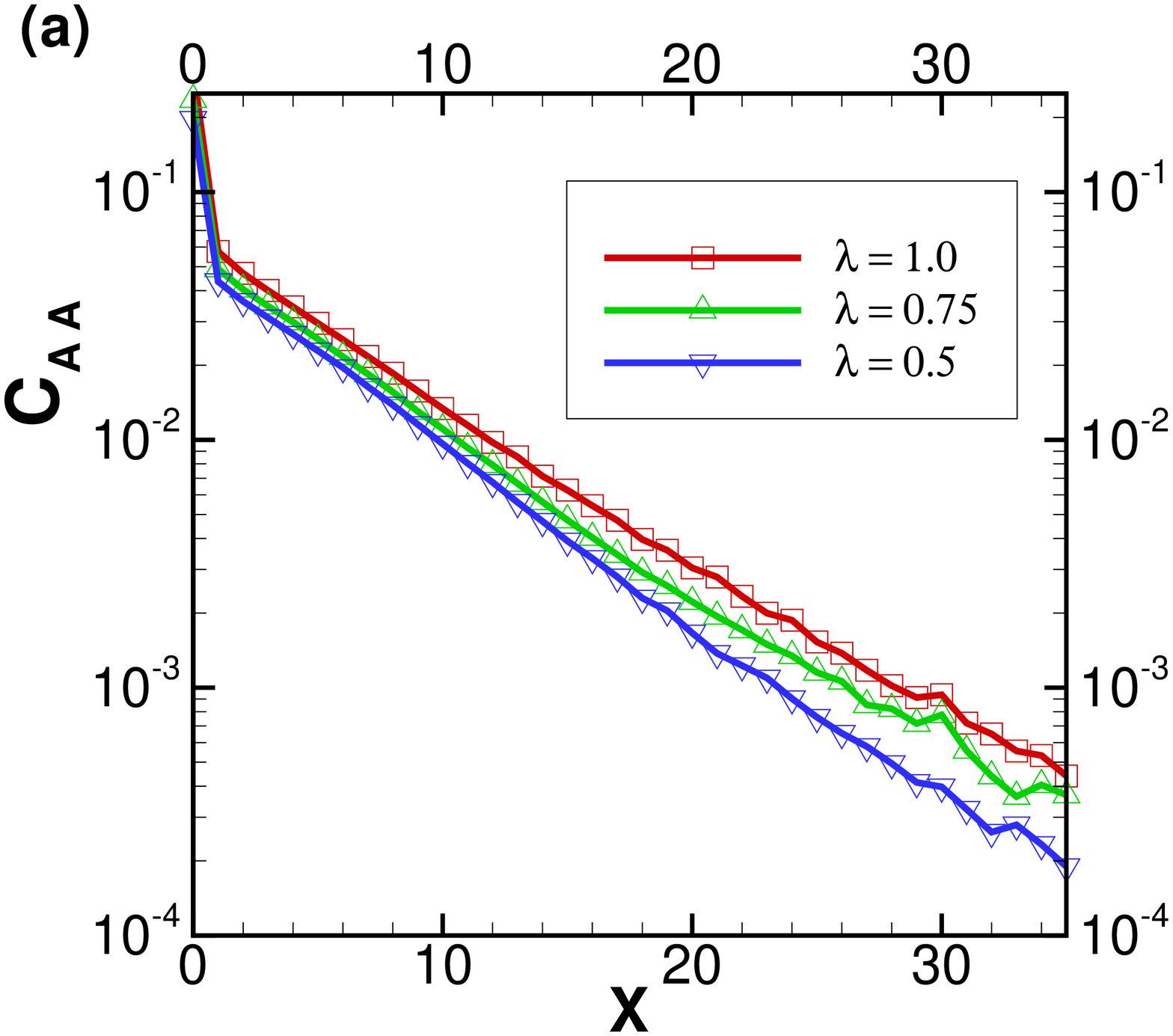} 
\includegraphics[width = 5.1cm]{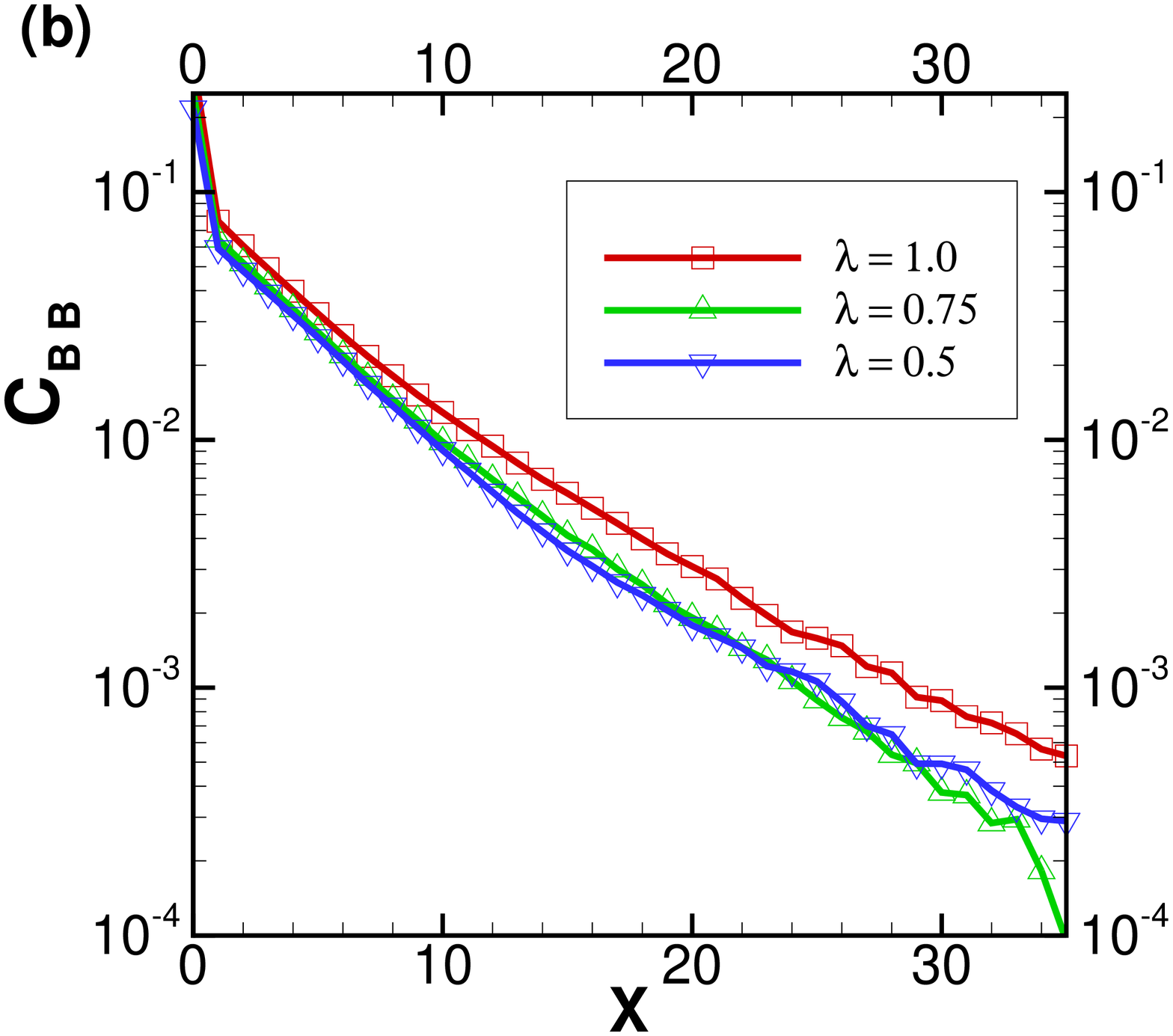} 
\includegraphics[width = 5.1cm]{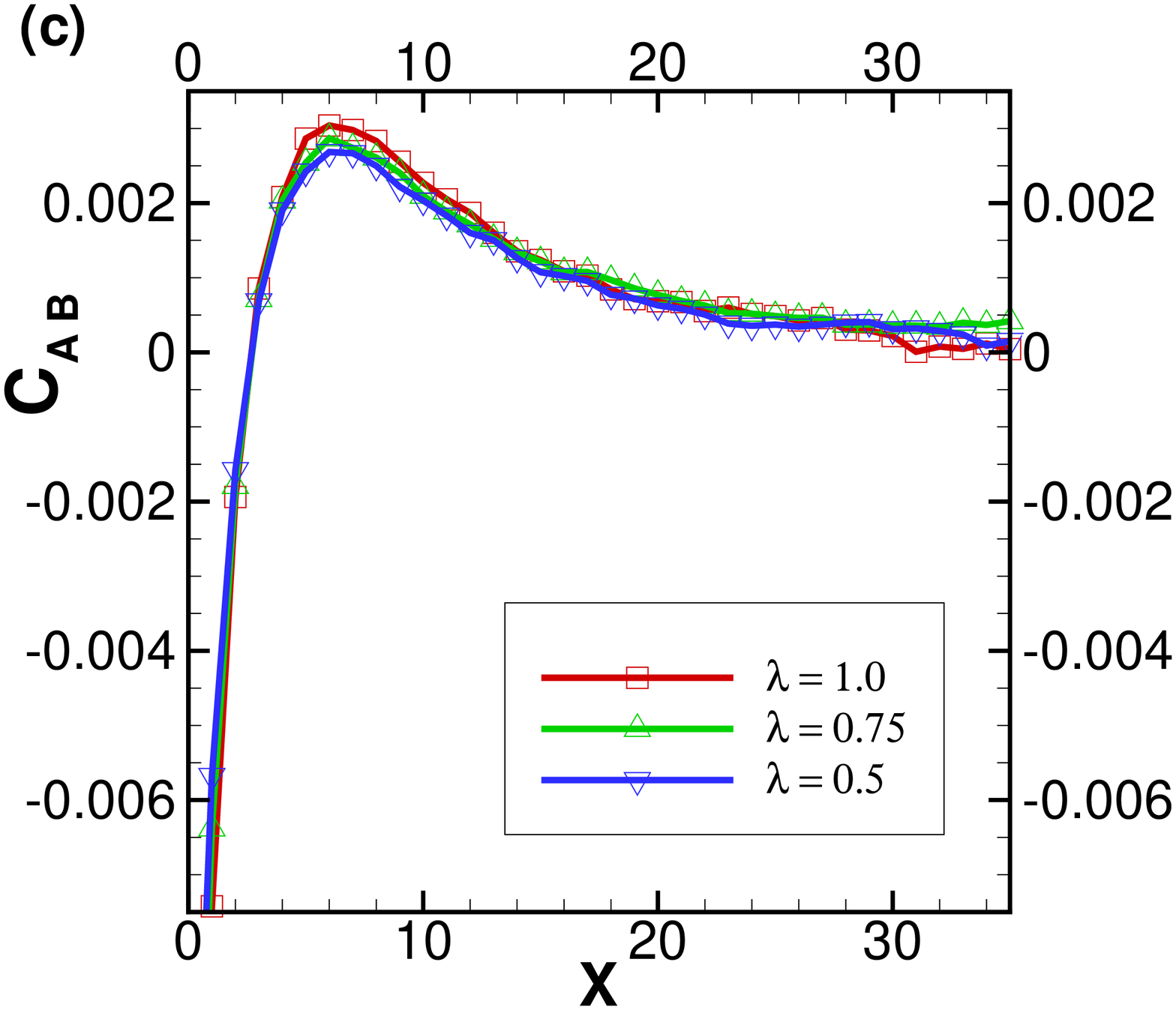}
\caption{\label{abcorr} 
        Static correlation functions (a) $C_{AA}(x)$, (b) $C_{BB}(x)$, and 
        (c) $C_{AB}(x)$, measured in simulations on a $1024 \times 1024$ 
	lattice, with rates $\sigma = 0.1$, $\mu = 0.1$, and with $\lambda$ 
	varied among $0.5$ (blue), $0.75$ (green), and $1.0$ (red). 
	(Colour online.)}
\end{center}
\end{figure}

The emerging spatial structures can be characterized quantitatively through the
static (and translationally invariant) correlation functions 
$C_{\alpha \beta}(x) = \langle n_\alpha(x) n_\beta(0) \rangle - 
\langle n_\alpha \rangle \, \langle n_\beta \rangle$, where 
$\alpha,\beta = A,B$, and $n_\alpha(x)$ denotes the occupation number for
particle species $\alpha$ at site $x$.
Figure~\ref{abcorr} depicts the results for $C_{AA}(x)$ (a), $C_{BB}(x)$ (b), 
and $C_{AB}(x)$ (c), as obtained from Monte Carlo simulations on a 
$1024 \times 1024$ square lattice, with rates $\sigma = 0.1$, $\mu = 0.1$, and 
various values of the predation rate $\lambda = 0.5$, $0.75$, and $1.0$.
Each set of measurements was initiated in a well-developed simulation, having 
run for $3000$ time steps, which was subsequently performed for an additional 
$1000$ time steps, while sampling the state of the system every $25$ steps. 
Thus, the correlation results are averaged over $40$ separate times throughout
the $1000$ additional steps.
The log-linear plots for $C_{AA}(x)$ and $C_{BB}(x)$ capture the essentially
exponentially decaying correlations of particles of the same species.
This suggests the form $C_{AA}(x) \propto C_{BB}(x) \approx F \, e^{-|x|/\xi}$,
with equal correlation lengths $\xi$ for the predators and prey; in addition,
both $\xi$ and the amplitude $F$ appear to depend only weakly on the predation
rate.
Qualitatively similar (but quantitatively different) to simulations with 
restriced site occupation numbers (compare figure~4 in \cite{Mobilia}), in
figure \ref{abcorr}(c) we observe local anti-correlations between the $A$ and
$B$ species for $0 \leq x \leq 2$ which are of course caused by the predation 
reaction.
There are pronounced positive correlations up to about $20$ lattice constants,
which indicates the width of the rather diffuse predator--prey activity fronts 
seen in figure~\ref{snapsh}.

\subsection{Population oscillations}

\begin{figure}[b] \begin{center}
\includegraphics[width = 7.7cm]{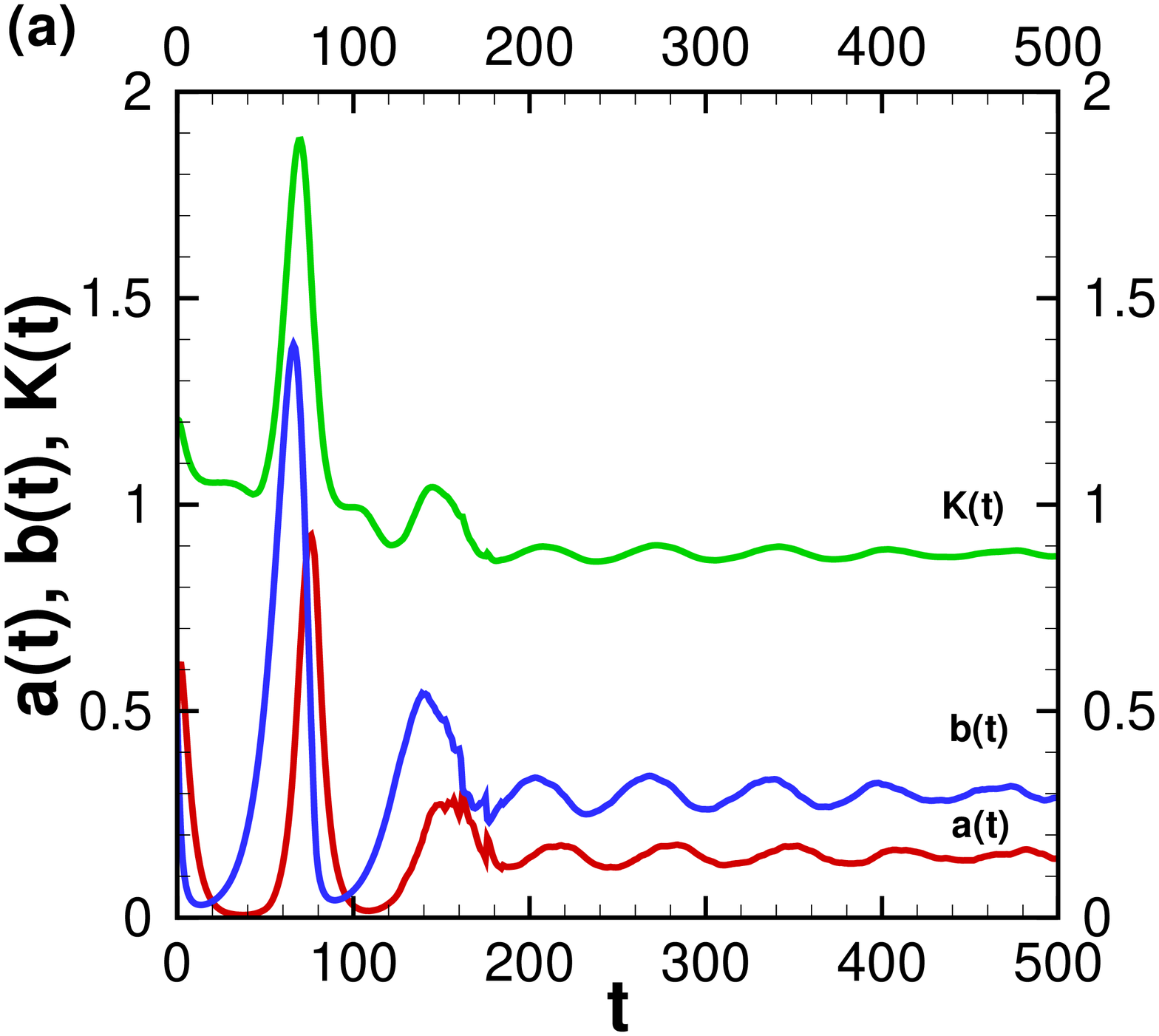} \
\includegraphics[width = 7.7cm]{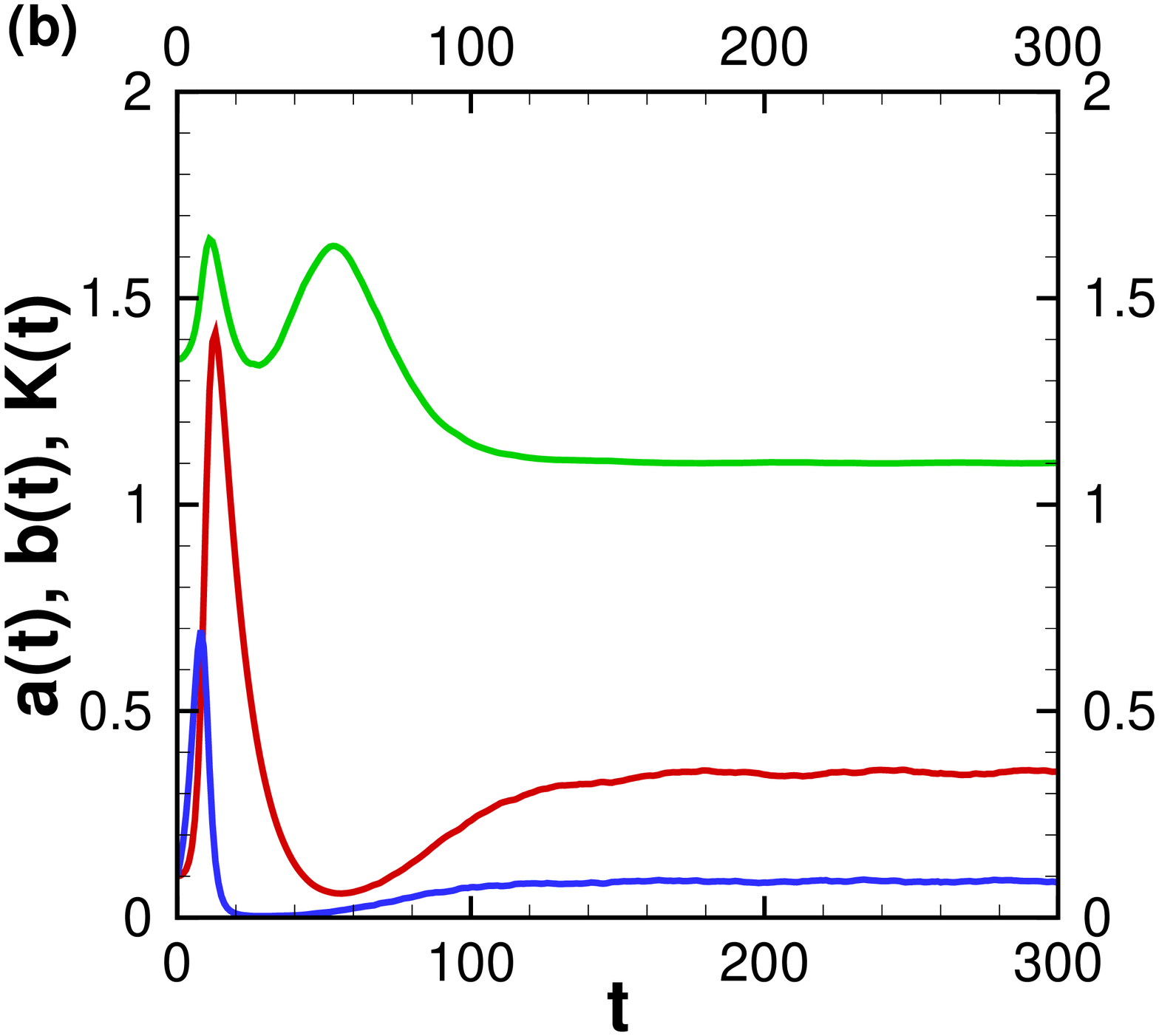}
\caption{\label{abkosc} 
	Early time evolution for the population density of predators $a(t)$ 
	(red), prey $b(t)$ (blue), and the quantity $K(t)$ (green) defined in 
	\eref{mfcons} from two single runs on a $1024 \times 1024$ lattice, 
	both starting with a random distribution with rates (a) $\sigma = 0.1$,
        $\mu = 0.2$, and $\lambda = 1.0$, and (b) $\sigma = 0.4$, $\mu = 0.1$, 
	and $\lambda = 1.0$. (Colour online.)}
\end{center}
\end{figure}

\begin{figure}[t] \begin{center}
\includegraphics[width = 7.7cm]{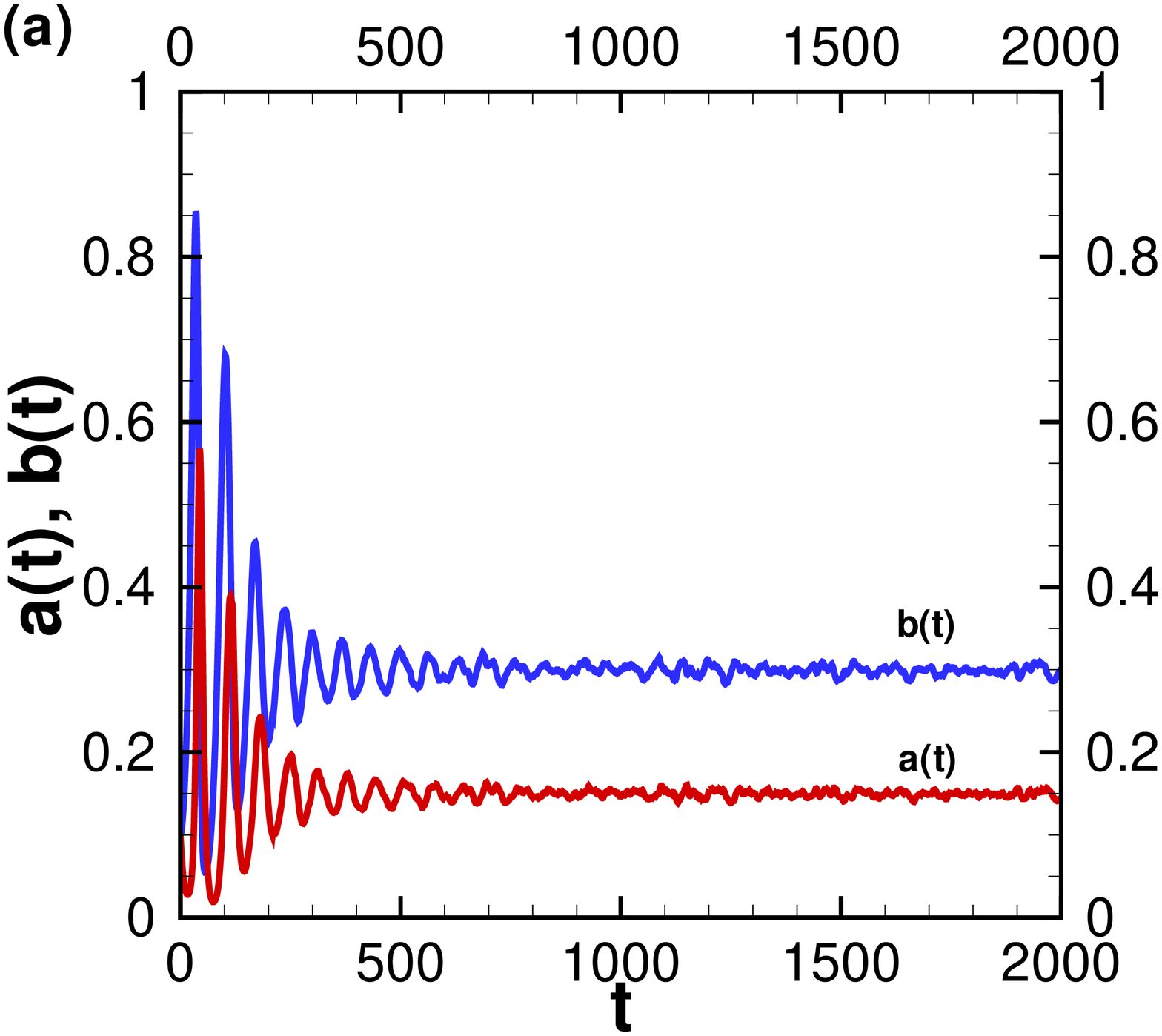} \
\includegraphics[width = 7.7cm]{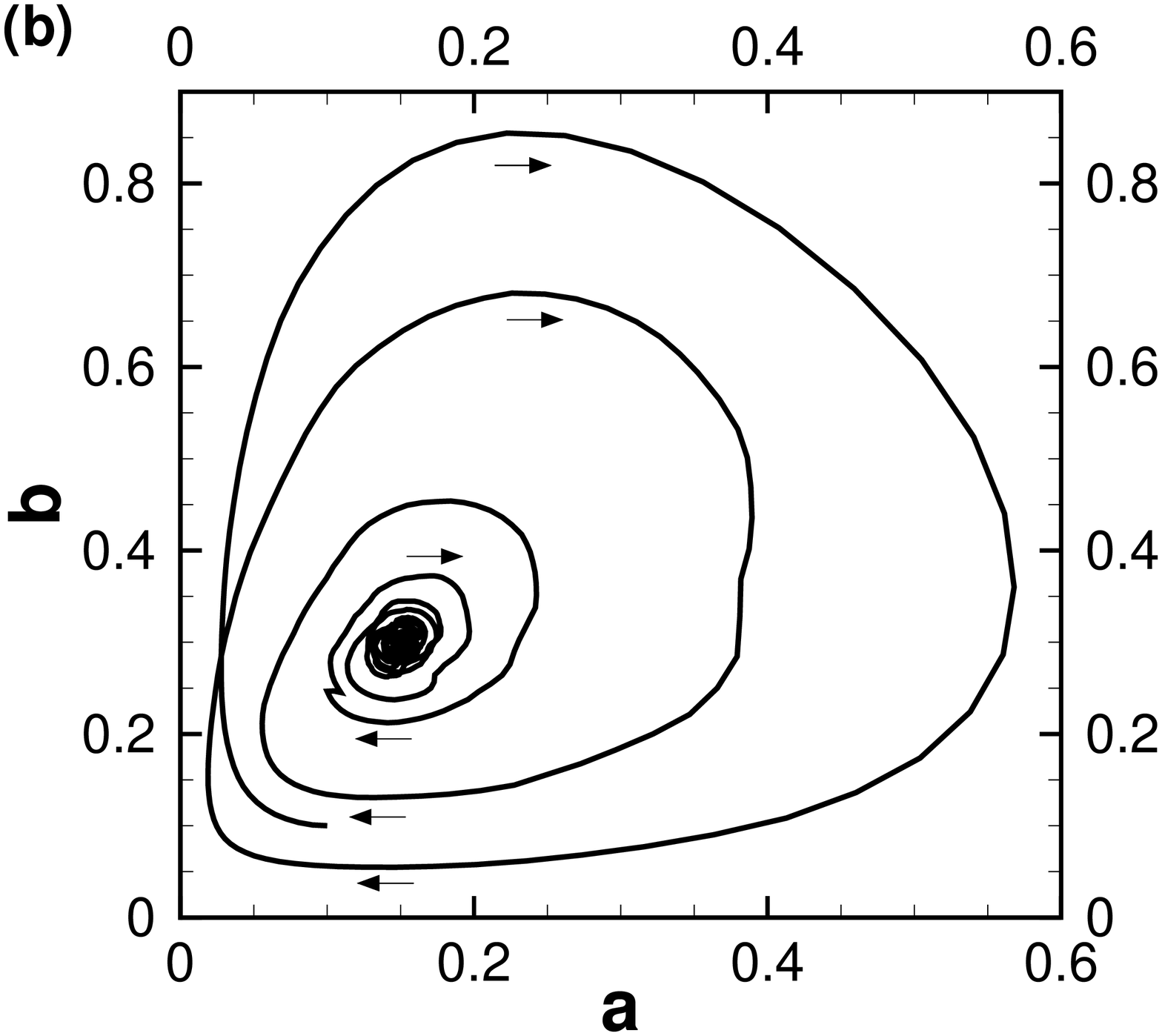} 
\caption{\label{oscill} 
	(a) Predators $a(t)$ (red) and prey $b(t)$ (blue) densities vs. time in
        a simulation run on a $1024 \times 1024$ lattice, with random initial 
	distribution, and rates $\sigma = 0.1$, $\mu = 0.2$, $\lambda = 1.0$, 
	and initial densities $a(0) = b(0) = 0.1$.
	(b) Trajectory in the $a$-$b$ plane from the simulation data shown in 
	(a), up to $t = 1000$. (Colour online.)}
\end{center}
\end{figure}

When the predator and prey total population densities $a(t)$ and $b(t)$ are 
plotted as functions of time $t$, one finds marked oscillations in the early 
time regime, as shown for two examples in figure~\ref{abkosc}.
These stochastic oscillations reflect the recurrent spatio-temporal structures 
visible in figure~\ref{snapsh}.
As time progresses, the amplitude of these transient fluctuations decreases
considerably, but as in simulations with site restrictions, in finite systems 
these transient oscillations may persist for a long time, see also 
figure~\ref{oscill}(a) below.
Increasing the reaction rates (with the hopping rate fixed) effectively renders
the processes more local, which suppresses front spreading and consequently 
reduces the amplitude of the population oscillations, as evident in 
figure~\ref{abkosc}(b).
In order to compare with the deterministic rate equations, we have also plotted
the quantity $K(t)$ defined in \eref{mfcons}, which according to the mean-field
approximation would remain constant.
Yet, we see that it in fact oscillates roughly in phase with the particle 
densities:
fluctuations evidently {\em amplify} at least the initial population cycles.
It is important to note that (in the thermodynamic limit) the asymptotic 
long-time mean population densities as well as $K(t \to \infty)$ are determined
by the reaction rates, and become {\em independent} of their initial values, as
is confirmed in figure~\ref{abkosc}.

As already apparent from the analysis of the corresponding rate equations, in
systems with limited prey carrying capacity an extinction threshold for the
predator population appears at sufficiently low values of the predation rate
$\lambda$ (with $\sigma$, $\mu$ held fixed).
We have investigated our lattice model with (almost) unlimited prey occupation
number in the limit of low values of $\lambda$, down to $\lambda \approx 0.02$ 
and found no signature of any phase transition: 
in the thermodynamic limit, this unconstrained predator--prey system seems to 
always allow species coexistence (as usual, however, {\em finite} systems 
should terminate in the absorbing state, albeit after potentially enormous 
crossover times).
Neither have we encountered a situation where the stable species coexistence
regime is governed by a stable node fixed point, which would be approached 
without any population oscillations; however, as mentioned before, for higher
reaction rates the oscillations become strongly damped.
Figure~\ref{oscill} depicts the population densities for an extended simulation
run with rates $\sigma = 0.1$, $\mu = 0.2$, $\lambda = 1.0$, and initial values
$a(0) = b(0) = 0.1$ (well away from the steady-state densities), along with the
corresponding phase space trajectory in the $a$-$b$ plane, which should be 
compared with the mean-field pictures in figure~\ref{mforbt}, computed with the
same rates.
Obviously the initial state determines only transient behaviour, the
long-time regime is governed by stochastic fluctuations about the attractive
fixed point in the centre of the graph.
We expect those to be quite well-described by the zero-dimensional effective
urn model and the `resonant amplification' mechanism described in 
\cite{McKane}.

\begin{figure}[t] \begin{center}
\includegraphics[width = 7.7cm]{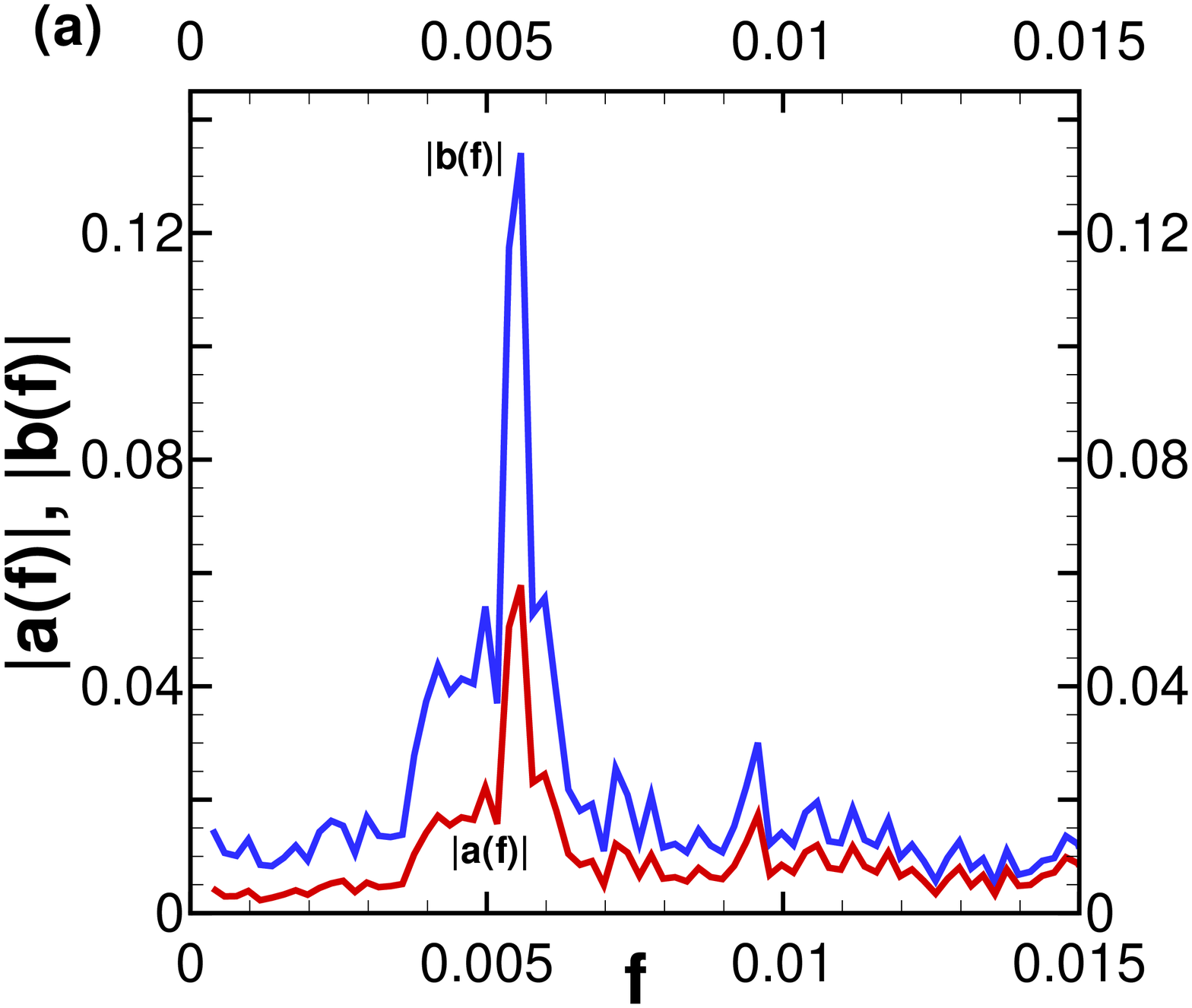} \
\includegraphics[width = 7.7cm]{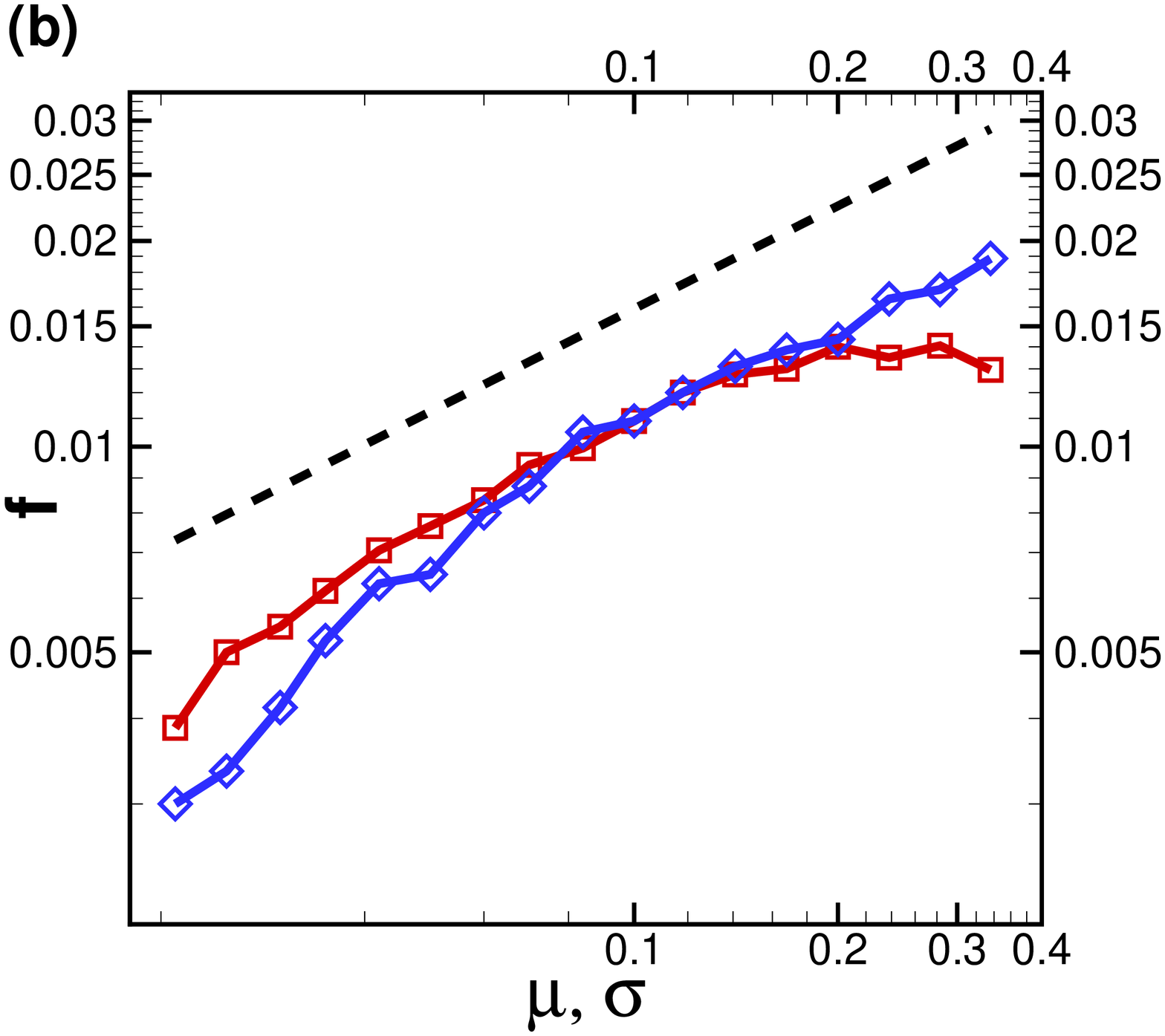}
\caption{\label{ftpeak} 	
	(a) Fourier transforms $|a(f)|$ and $|b(f)|$ of the predator (red) and 
	prey (blue) density data for a simulation run on a $1024 \times 1024$
	lattice with rates $\sigma = 0.03$, $\mu = 0.1$, and $\lambda = 1.0$, 
	as function of frequency $f$.
	(b) Variation of the characteristic peak frequencies in $|a(f)|$ and 
	$|b(f)|$ with $\sigma$ (red squares) and $\mu$ (blue diamonds), with 
	the respective other rate held fixed at the value $0.1$ and 
	$\lambda = 1.0$, as obtained from simulation data on $1024 \times 1024$
	lattices up to time $t = 20000$, compared with the result 
	$f = \sqrt{\mu \, \sigma} / 2 \pi$ from (linearized) mean-field theory
	(black dashed). (Colour online.)} 
\end{center}
\end{figure}

Additional information can be extracted from the simulation data by studying
the (fast) Fourier transforms $a(f) = \int a(t) \, \rme^{2 \pi \rmi f t} \, dt$
and similarly $b(f)$ of the time signals $a(t)$ and $b(t)$.
As shown in figure~\ref{ftpeak} for runs with $\sigma = 0.03$, $\mu = 0.1$, and
$\lambda = 1.0$, extended to $t = 20000$, the predator and prey Fourier signals
display prominent peaks at the same characteristic frequency, which evidently 
governs the erratic oscillations.
In order to assess the validity of the linearized mean-field approximation
result $f = \sqrt{\mu \, \sigma} / 2 \pi$ quantitatively, we have obtained the
peak frequency values for various Monte Carlo simulations on $1024 \times 1024$
square lattices, all run with $\lambda = 1.0$ and with either $\mu$ or $\sigma$
held fixed at the value $0.1$, while the respective other rate varied between 
$0.02$ and $0.36$.
The results are displayed as a double-logarithmic plot in 
figure~\ref{ftpeak}(b).
Interestingly, the simulation data show a more pronounced deviation from the 
square-root dependence on $\mu$ than on $\sigma$ for low rates, while the
reverse is true for large rates.
Moreover, the characteristic peak frequencies are reduced by about $50 \%$ as 
compared with the mean-field prediction, obviously renormalized by stochastic 
fluctuations; similar effects are found in site-restricted simulations (see 
figure~9 in \cite{Mobilia}).
It is remarkable though that for values between $0.04$ and $0.24$ the curves 
for varying rates $\mu$ or $\sigma$ appear to coincide, which would indicate 
the simple functional dependence $f(\mu,\sigma) = {\tilde f}(\mu \, \sigma)$ in
this range.

\subsection{Age distributions}

\begin{figure}[t] \begin{center}
\includegraphics[width = 7.7cm]{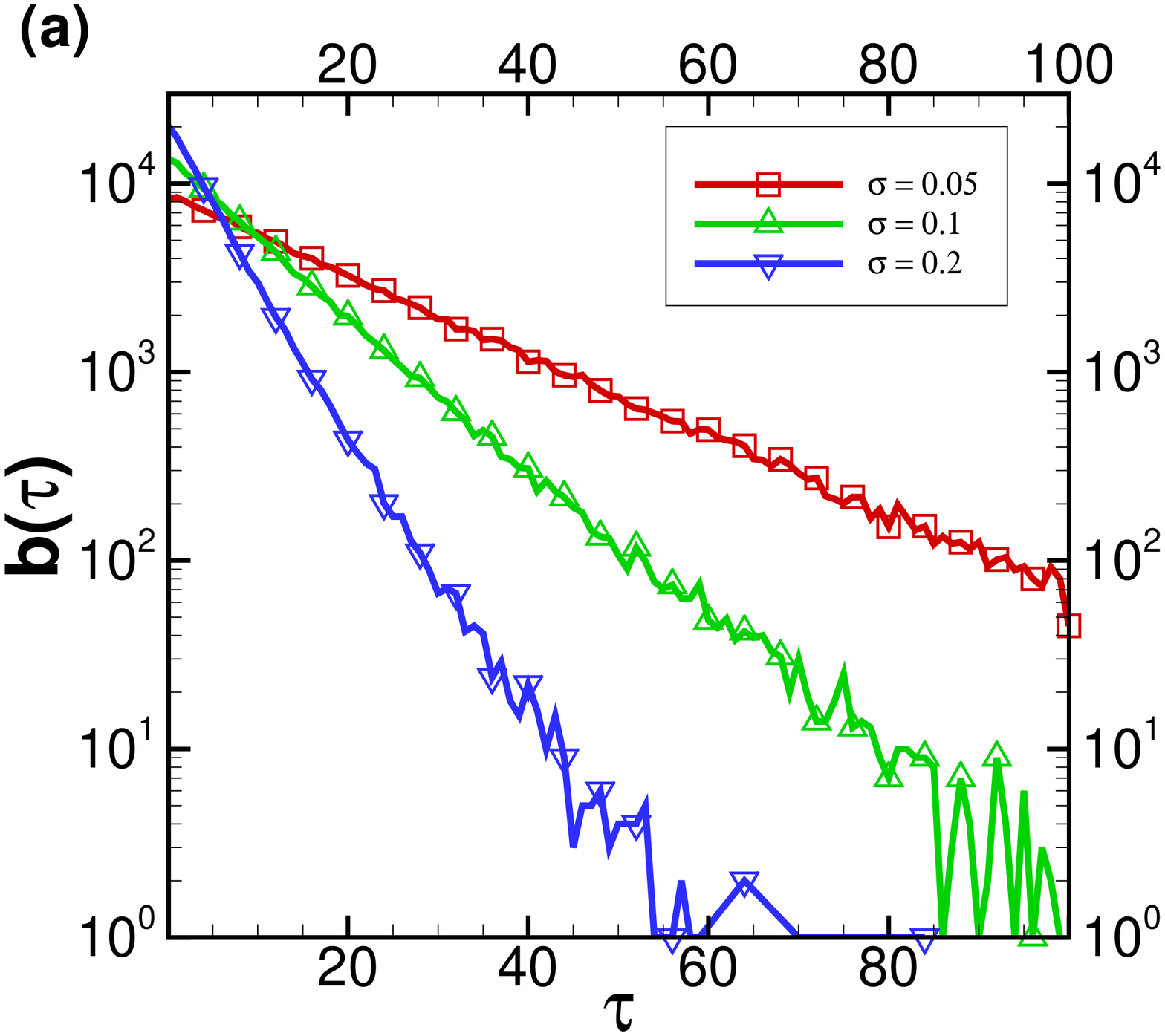} \
\includegraphics[width = 7.7cm]{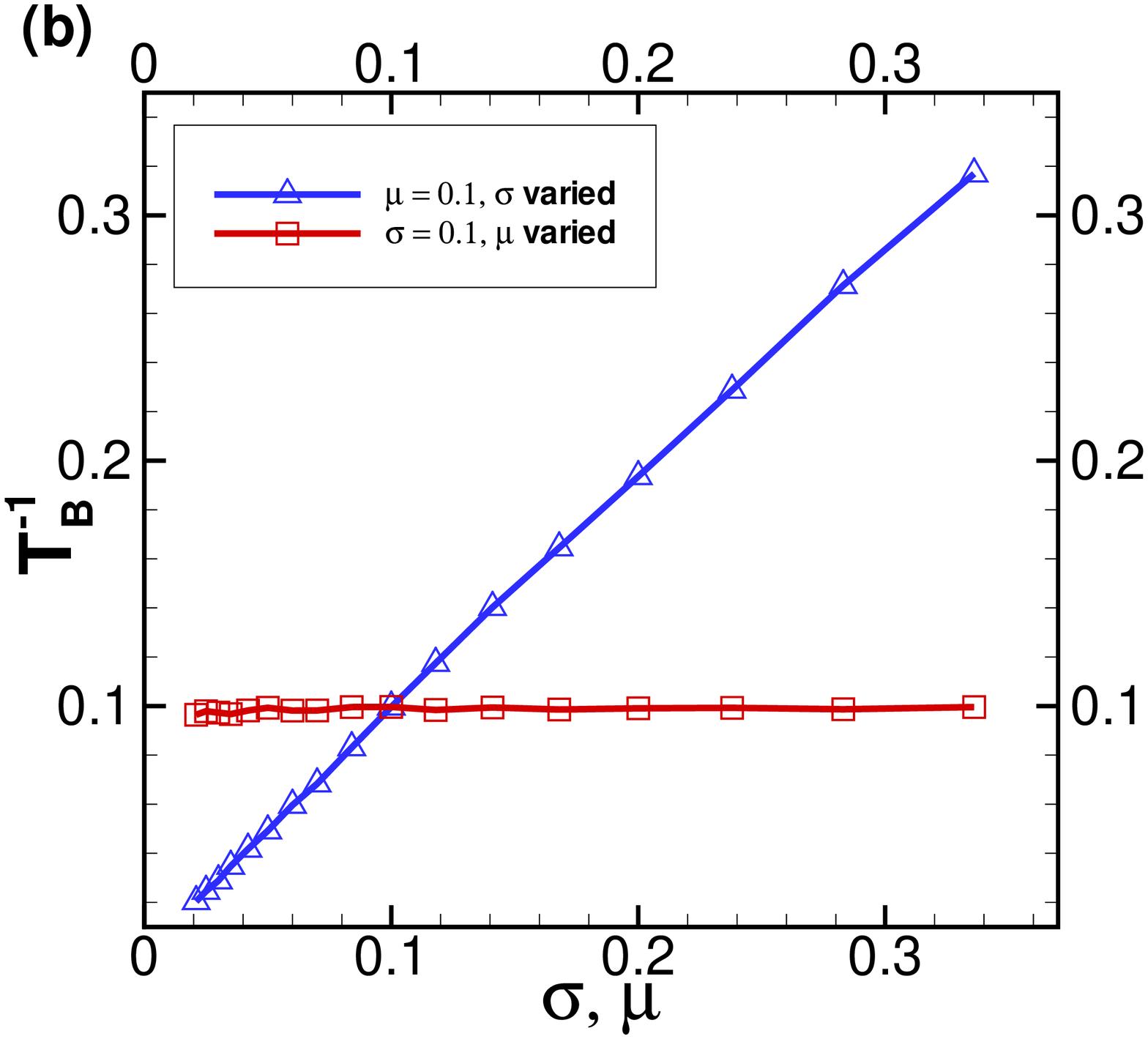} 
\caption{\label{bagdis} 
	(a) Prey age distribution histogram $b(\tau)$ as function of their
        life time $\tau$, obtained from Monte Carlo simulations on a 
	$1024 \times 1024$ lattice with $\mu = 0.1$, $\lambda = 1.0$, and
	$\sigma = 0.05$ (red), $0.1$ (green), and $0.2$ (blue).
	(b) Inverse mean prey life times $T_B^{-1}$ as measured in similar
	simulations, with fixed $\lambda = 1.0$ and either $\sigma$ (blue) or 
	$\mu$ (red) varying, and the respective other rate held constant at 
	$0.1$. (Colour online.)}
\end{center}
\end{figure}

We have also obtained the age histograms $b(\tau)$ for the total number of 
surviving prey up to time $\tau$ after they were produced.
The log-linear plot in figure~\ref{bagdis}(a), taken for $\mu = 0.1$, 
$\lambda = 1.0$, and different values of $\sigma = 0.05, 0.1, 0.2$ on a 
$1024 \times 1024$ square lattice, shows that the prey age distribution 
$b(\tau)$ essentially decays exponentially with $\tau$.
From the slopes in these graphs we can read off the prey population inverse
mean life time $T_B^{-1}$, which is plotted in figure~\ref{bagdis}(b) as 
function the rates $\sigma$ (with $\mu = 0.1$ and $\lambda = 1.0$ held fixed), 
and as function of $\mu$ ($\sigma = 0.1$, $\lambda = 1.0$).
One finds that $T_B^{-1}(\sigma,\mu) \approx \sigma$ for the rate interval 
studied here, whereas it displays no markedly significant dependence on $\mu$.
This is indeed borne out already upon linearizing the mean-field rate equations
\eref{lvreqa} about the steady state 
$(a_u = \sigma/\lambda , b_u = \mu/\lambda)$, which yields an effective prey
death rate $\sigma$ induced by the predation reaction, independent of $\mu$.

\subsection{One-dimensional simulations}

\begin{figure}[t] \begin{center}
\includegraphics[width = 8.2cm]{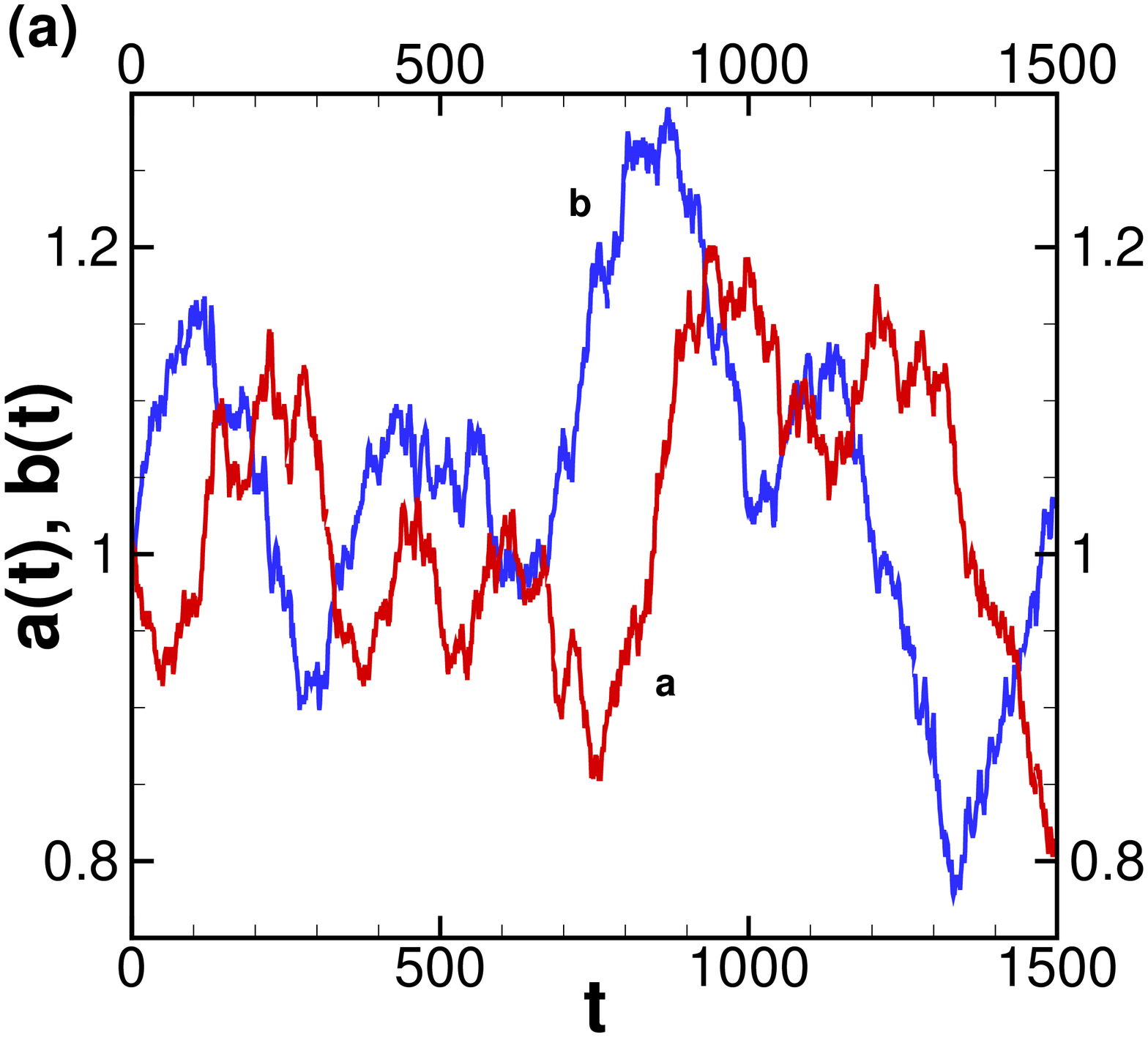} \
\includegraphics[width = 6.8cm]{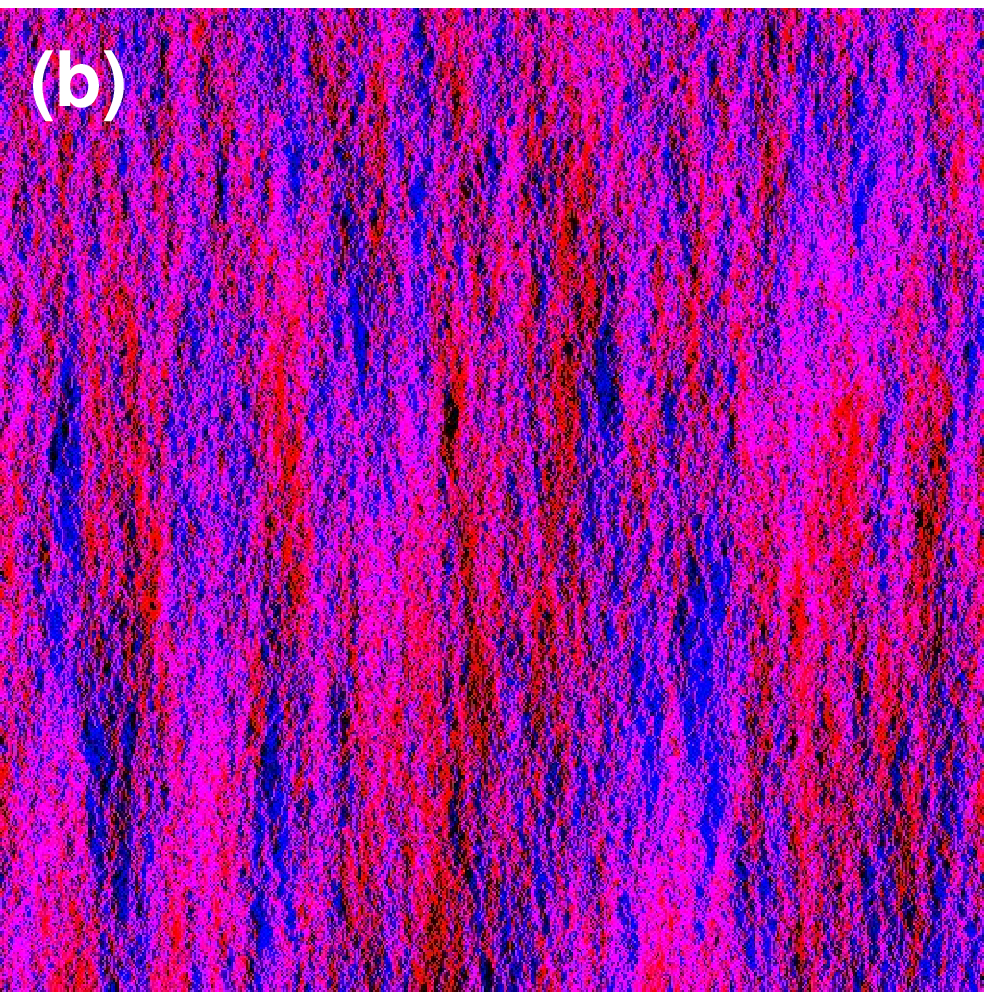} 
\caption{\label{1ddiff} 
        (a) Predator (red) and prey (blue) population densities in a (single) 
        one-dimensional simulation on $512$ lattice sites with rates 
	$\sigma = 0.01$, $\mu = 0.01$, $\lambda = 0.01$, and $a(0) = b(0) = 1$.
	(b) Space-time plot, with time running from top to bottom (up to 
	$t = 500$) for this diffusion-dominated run, where red and blue pixels 
	respectively indicate sites with at least one predator ($A$) and prey 
	($B$); magenta sites are occupied by at least one representative of 
	either species; black sites are empty. (Colour online.)}
\end{center}
\end{figure}

\begin{figure}[b] \begin{center}
\includegraphics[width = 8.2cm]{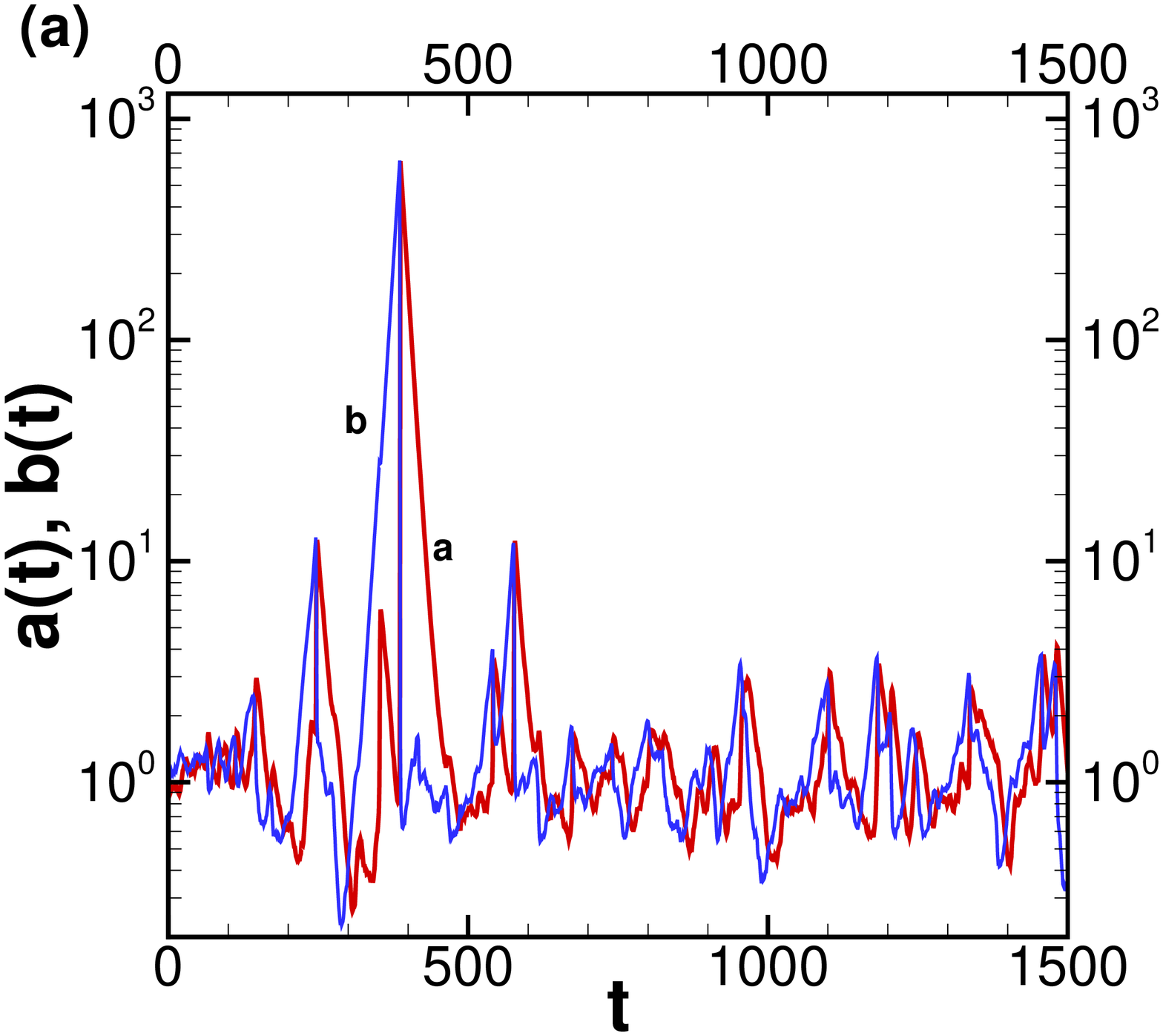} \
\includegraphics[width = 6.8cm]{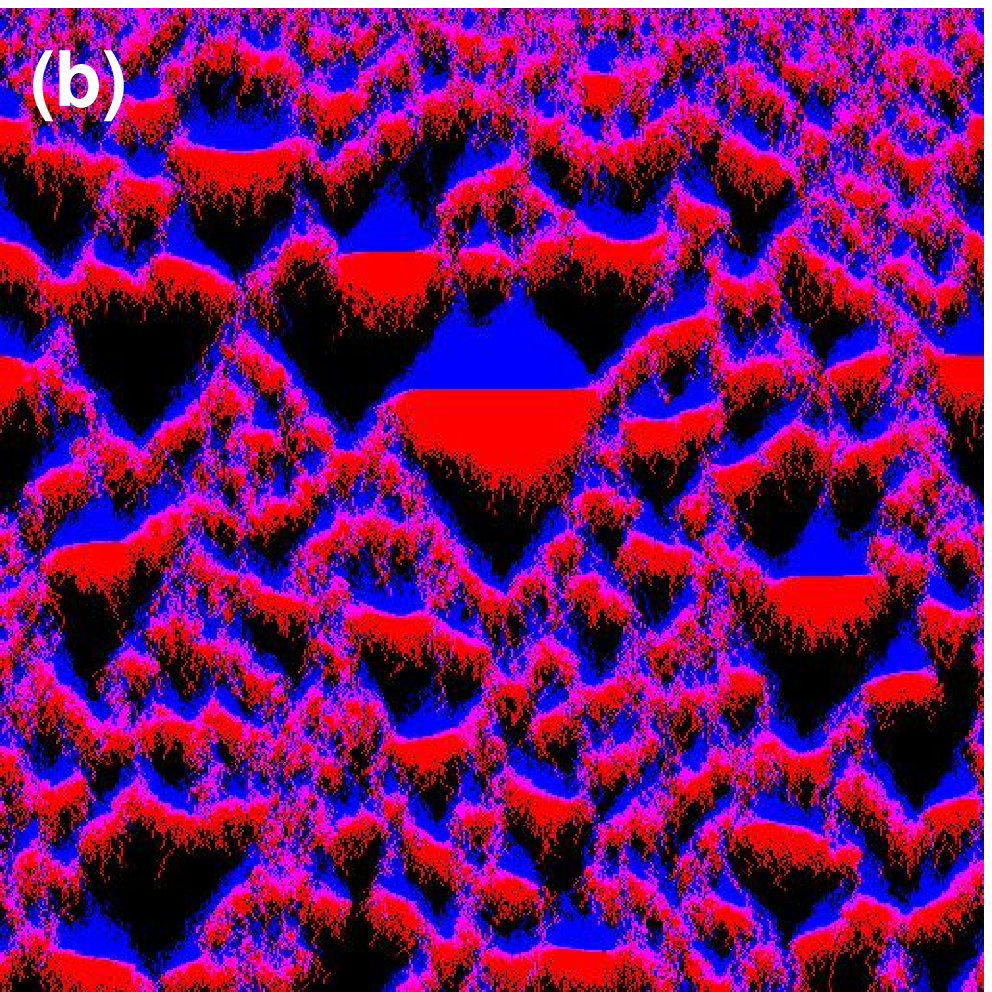} 
\caption{\label{1dreac} 
	(a) Predator (red) and prey (blue) population densities in a (single)
        one-dimensional simulation on $512$ lattice sites with rates 
	$\sigma = 0.1$, $\mu = 0.1$, $\lambda = 0.1$, and $a(0) = b(0) = 1$, 
	and corresponding space-time plot, with time running from top to 
	bottom (up to $t = 500$) for this reaction-dominated run.
	(Colour online.)}
\end{center}
\end{figure}

We have also run Monte Carlo simulations for our unconstrained lattice 
Lotka--Volterra model in one dimension.
The total population densities as well as space-time plots for the first $500$ 
time steps are shown for two representative examples in figures~\ref{1ddiff}
and \ref{1dreac}.
In the first run, the reaction rates are low (all $0.01$), and the kinetics is
dominated the diffusion.
Correspondingly, the system remains well mixed, and displays a stochastic time
signal.
In the second simulation, with all rates set to $1.0$, the on-site reactions
dominate. 
Localized prey population bursts occur, but invading predators quickly remove 
emerging prey clusters, subsequently proliferate and then die out, which leads 
to intriguing wedge-like structures in the early time regime.
At any rate, we generally observe species coexistence during the duration of
our runs, in accord with earlier investigations of a one-dimensional four-state
system \cite{Lipowska}.
This is in stark contrast with simulations with restricted site occupation 
numbers, where the system tends towards eventual predator extinction, with the 
asymptotic approach to the absorbing state proceeding via very slow coarsening 
processes, namely merging of the active domains, presumably governed by the
$t^{-1/2}$ power law of single-species coagulation (see figure~11 in 
\cite{Mobilia}).

\subsection{Simulations with restricted predation}

As mentioned in section~3.2, we have also performed simulations wherein an $A$
particle in each of its moves to a new site could annihilate at most one $B$ 
particle there.
In effect, this sets an upper limit to the efficiency of the predation process,
at variance with the rate equations \eref{lvreqa}, which induces interesting 
differences to the previous simulations with multiple simultaneous predation 
events when the prey density becomes large.
At large predation rates $\lambda > \mu$, the qualitative behaviour of this
model variant is essentially as described above.
However, there emerges one distinction when the reaction rates are large (with 
the hopping rate $D$ held fixed), and the on-site reactive processes dominate.
In this situation, instead of spreading diffuse predator--prey fronts, we have
observed pulsating activity zones.
In addition to the familiar erratic oscillations, these induce intermittent 
population spikes that tend to dominate the long-time properties of the system.
(In fact, these localized population explosions set limits to the rates we 
could explore, because the maximum site occupation number bounds may become
exceeded.)

More dramatically, when $\lambda$ approaches $\mu$ from above, predation 
becomes too inefficient for the predators, and we find an extinction threshold 
for the $A$ population at (or very near) $\lambda \approx \mu$.
For $\lambda < \mu$, $a(t)$ appears to decay exponentially; at the transition,
however, the predator density can apparently assume any value.
A proper characterization of this unusual extinction transition would however 
require considerable additional effort.

\section{Discussion and Conclusion}

Understanding biodiversity and identifying mechanisms allowing to maintain 
coevolution, as well as the influence of spatial distribution of the agents, 
are central problems in modern theoretical biology and ecology. 
In this context, we have studied a stochastic lattice version of the
Lotka--Volterra model for the dynamics of two competing populations.
As a main difference with numerous earlier studies, we have performed Monte
Carlo simulations without any site restrictions, which can be interpreted in 
the biological or ecological context as a system without local limitation of 
the growth rate. 
This study is also of interest from a physical viewpoint because, at the 
determinsitic mean-field rate equations level, one naturally recovers the 
genuine Lotka--Volterra equations from the present stochastic model system. 
This investigation also allows to test further the robustness of stochastic 
predator--prey systems, which have been recently shown to share numerous 
properties. 
From a biological and ecological perspective, it is relevant to understand 
better the role of the presence or absence of some form of spatial limitation 
of the resources, which can be (arguably) simply mimicked by considering site 
restricted and unrestricted stochastic models, respectively.

After having briefly outlined the basic features of the Lotka--Volterra model
in the framework of the mean-field rate equations, both for infinite and finite
carrying capacity of the prey population, and explaining the procedure we have
developed to efficiently structure the data, we have reported results of 
extensive simulations of our unrestricted stochastic predator--prey system.
As a major difference with respect to the results for site-restricted models, 
we have found no evidence of any extinction threshold in our one- and 
two-dimensional simulations.
With both types of agents being allowed to locally proliferate, both species 
are always found to coexist.
Similar to the site-restricted models deep in the coexistence phase, our 
unrestricted system is characterized by complex and correlated patterns 
emerging from `pursuit and evasion' wave fronts. 
These are rendered more diffuse in the present unrestricted simulations through
regions accomodating both predators and prey on the same spots.
The quantitative properties (characteristic correlation lenghts) of the 
patterns have been studied by computing various static correlation functions.
The dynamical properties have been investigated by considering both time 
dependence of the densities and the trajectories in the phase portrait. 
As for restricted stochastic models, the population densities are typically 
characterized by damped erratic oscillations, with an amplitude that vanishes 
in the thermodynamic limit. 
However, as a novel feature, when the reaction rates dominate over the 
diffusive process, the reaction kinetics takes place largely locally (on-site),
and one observes, depending on the model variant, either quickly decaying
oscillations, or pulsating activity zones and spikes in the density profiles. 
By means of a Fourier analysis, and similar to the restricted case, we have 
found that the characteristic frequency of the damped and fluctuation-induced 
oscillations is markedly reduced with respect to the mean-field prediction, 
while its functional dependence seems in reasonable agreement with the 
deterministic result.
We have also studied the prey age distribution and shown that it decays 
exponentially, with an inverse mean-life time independent of the death rate of 
the predators and depending linearly on the reproduction rate of the prey.

\ack
This work was in part supported by U.S. National Science Foundation through
grant NSF DMR-0308548.
MM gratefully acknowledges the support from the German Alexander von Humboldt 
Foundation through Fellowship No. IV-SCZ/1119205 STP.
We would like to thank E.~Frey, I.T.~Georgiev, R.~Kulkarni, A.~McKane, 
T.~Newman, S.~Redner, T.~Reichenbach, B.~Schmittmann, N.~Shnerb, and 
R.~Stinchcombe for inspiring discussions.

\end{document}